\documentclass[journal]{IEEEtran}
\usepackage[T1]{fontenc}

\usepackage{amsmath,amssymb}
\usepackage{graphicx,epsfig,color,verbatim}

\newtheorem{thm}{Theorem}[section]

\newtheorem{prop}[thm]{Proposition}

\newtheorem{theorem}{Theorem}[section]

\newtheorem{definition}[theorem]{Definition}
\newtheorem{proposition}[theorem]{Proposition}
\newtheorem{corollary}[theorem]{Corollary}

\newcommand{\mb}{\mathbf}

\newcommand{\bs}{\boldsymbol}
\newcommand{\mc}{\mathcal}
\newcommand{\bd}{\textbf}

\newcommand{\ds}{\displaystyle}

\newcommand{\mt}{\mathtt}

\begin{document}

\title{Green Power Control in Cognitive Wireless Networks}

\author{
\authorblockN{
Mael Le Treust\authorrefmark{1} and Samson Lasaulce\authorrefmark{2} and Yezekael Hayel
\authorrefmark{3} and Gaoning He\authorrefmark{4}}

\thanks{Part of this paper has been published in Infocom 2011. This work is also partially supported by L'Agence Nationale de la Recherche (ANR) within the project ANR-09-VERS0: ECOSCELLS.}

\authorblockA{\authorrefmark{1}
Universit\'e INRS, Centre Energie, Mat\'eriaux et T\'el\'ecom, 800 rue de la Gaucheti\`ere, 
Montr\'eal, Canada }

\authorblockA{\authorrefmark{2}
Laboratoire des Signaux et Syst\`emes, Sup\'elec,
3 rue Joliot Curie, 91191 Gif-sur-Yvette, France}

\authorblockA{\authorrefmark{3}
LIA/CERI, University of Avignon,  Agroparc BP 1228, 84911 Avignon Cedex 9, France }\\

\authorblockA{\authorrefmark{4}
HUAWEI, Central Research Institute, 2222 Xinjinqiao Rd, Pudong, 201206, Shanghai, China }\\
}

\maketitle

\begin{abstract}
A decentralized network of cognitive and non-cognitive transmitters where each transmitter aims at maximizing his energy-efficiency is considered. The cognitive transmitters are assumed to be able to sense the transmit power of their non-cognitive counterparts and the former have a cost for sensing. The Stackelberg equilibrium analysis of this $2-$level hierarchical game is conducted, which allows us to better understand the effects of cognition on energy-efficiency. In particular, it is proven that the network energy-efficiency is maximized when only a given fraction of terminals are cognitive. Then, we study a sensing game where all the transmitters are assumed to take the decision whether to sense (namely to be cognitive) or not. This game is shown to be a weighted potential game and its set of equilibria is studied. Playing the sensing game in a first phase (e.g., of a time-slot) and then playing the power control game is shown to be more efficient individually for all transmitters than playing a game where a transmitter would jointly optimize whether to sense and his power level, showing the existence of a kind of Braess paradox. The derived results are illustrated by numerical results and provide some insights on how to deploy cognitive radios in heterogeneous networks in terms of sensing capabilities.
\end{abstract}


\section{Introduction}


In fixed communication
 networks, the paradigm of peer-to-peer
communications has known a powerful surge of interest during the
past two decades with applications such as the Internet. Remarkably,
this paradigm has also been found to be very useful for wireless
networks. Wireless ad hoc and cognitive networks are two illustrative
examples of this. One important typical feature of these networks is
that the terminals have to take some decisions in an autonomous
or quasi-autonomous manner. Typically, they can choose their
power control and resource allocation policy. The corresponding
framework, which is the one of this paper, is the one of decentralized or distributed
power control (PC) or resource allocation. More specifically, the
scenario of interest is the case of power control over
quasi-static channels in cognitive
networks \cite{Fette}. In such a context, which is broader than the one of ad hoc and cognitive wireless networks, we assume that some (possibly all) transmitters are able to sense the power levels of non-cognitive transmitters and adapt their power level
accordingly. The considered model of multiuser networks is a multiple
access channel (MAC) with time-selective non-frequency selective links but the methodology can be applied to other types of interference networks. Technical issues related to spectrum usage is not
considered in this paper, leaving this aspect as a relevant extension of this paper. Rather, we want to study the effect of cognition in terms of energy usage, the potential benefits in terms of spectral efficiency having been well investigated. The selected performance metric for a transmitter is derived from the
energy-efficiency definition of \cite{goodman-pc-2000}. The authors of \cite{goodman-pc-2000} define energy-efficiency as the number of bits successfully decoded by the receiver per joule consumed at the transmitter (in \cite{goodman-pc-2000} the radiated power is concerned). More specifically, the authors analyze the problem of decentralized
power control in flat fading multiple access channels. The problem is formulated as a non-cooperative one-shot
game where the players are the transmitters, the action of
a given player is her/his/its transmit power (''his'' is chosen
in this paper), and his payoff/reward/utility function is the
energy-efficiency of his communication with the receiver; we will not provide here the motivations for using game theory to study distributed power control problems but some of them can be found e.g., in \cite{lasaulce-book-2011}. The
results reported in \cite{goodman-pc-2000} have been extended to the case
of multi-carrier systems in \cite{meshkati-jsac-2006}.

The framework of the present work is close in spirit to \cite{goodman-pc-2000,meshkati-jsac-2006} but differs from them in several aspects. The most important one is that there can be a hierarchy among the transmitters in terms of observation capabilities, some transmitters can be cognitive and observe the others whereas the latter cannot observe the actions played by the former. Technically, this leads to a Stackelberg-type formulation of the problem \cite{stackelberg-book-1952}. The closest work to the one reported here is \cite{lasaulce-twc-2009} where a Stackelberg model of energy-efficient power control problems is introduced for the first time. The present work reports a significant extension of the framework introduced in \cite{lasaulce-twc-2009}. Two games are studied in detail. The power control game corresponds to a generalization of the game addressed in \cite{lasaulce-twc-2009}~: the sensing costs are taken into account (observing/sensing the others has a cost) and more importantly, our analysis is not limited to one non-cognitive transmitter (i.e., a single game leader). Then, we introduce a new game where the transmitters decide whether to sense or not. A third game, which is is an hybrid control game and include the two mentioned games as special cases, is shown to be not worth being studied because of the existence of a Braess paradox \cite{Braess-69}.


The paper is organized as follows. In Sec. \ref{problem}, the assumed signal model to describe the distributed power control problem over time-selective non-frequency selective multiple access channels is provided. Known results concerning the case where the transmitters tune their power levels from block to block in a distributed way and without observing the other transmitters (i.e., they cannot sense the powers chosen by the others) are provided. In Sec. \ref{sec:power-control-game}, we assume that some transmitters have sensing capabilities, which creates a hierarchy in terms of observation capabilities between the transmitters. The effect of this is that choosing rational power control policies in this setting leads to a more efficient network outcome (a Stackelberg equilibrium), provided that the sensing cost for a cognitive transmitter is not too high. While in Sec. \ref{sec:power-control-game}, a transmitter was imposed to sense or not, it is assumed in Sec. \ref{sec:sensing-game} that this choice is left to the transmitter itself. It is shown that there exists an optimal number of cognitive transmitters in terms of network utility and therefore having too many advanced terminals can be detrimental to the global performance. It is shown that leaving the choice to a transmitter to choose in a joint manner its power level and whether to sense is in fact less energy-efficient than imposing that the transmitters choose these two quantities separately. This shows the interest in studying the power control game (as in Sec. \ref{sec:power-control-game}) and the sensing game separately. The sensing game is a new game we introduce and is shown to possess attractive properties for distributed optimization and learning algorithms. Finally, in Sec. \ref{sec:numerical-results} numerical illustrations are provided and the paper is concluded in Sec. \ref{Conclusion}.

\section{Problem statement} \label{problem}

\subsection{System model}
\label{sec:signal-model}

We consider a decentralized multiple access channel with a finite number of
transmitters, which is denoted by $K$. The network is said to be
decentralized in the sense that the receiver (e.g., a base/mobile station)
does not dictate to the transmitters (e.g., mobile/base stations) their
power control policy. Rather, all the transmitters choose their policy by
themselves and want to selfishly maximize their energy-efficiency. In particular, they can ignore some specified centralized policies. We assume that the users transmit their data over time-selective non-frequency selective channels, at the same time and on the same frequency band; channels are considered to be constant over each block of data. Note that a block
is defined as a sequence of $M$ consecutive symbols which comprises
a training sequence that is, a certain number of consecutive symbols
used to estimate the channel (or other related quantities)
associated with a given block. A block has therefore a duration less
than the channel coherence time. The equivalent baseband signal at the receiver can be written as
\begin{equation}
\label{eq:received-signal}
 y(t) = \sum_{k=1}^{K} h_k x_k(t) + z(t)
\end{equation}
where $k  \in \mc{K}$, $\mc{K} = \{1,...,K\}$, $x_k(t)$ represents
the symbol transmitted by transmitter $k$ at time $t\in\mathbb{N}$,
$\mathbb{E}|x_k|^2 = p_i$, the noise $z$ is assumed to be
decentralized according to a zero-mean Gaussian random variable with
variance $\sigma^2$ and each channel gain $h_k$ varies over time but
is assumed to be constant over each block~; the symbol index $t$ will be omitted in this paper. In terms of channel state information (CSI), the receiver is assumed to know all the channel gains (global CSI) while each transmitter only knows his own channel (local CSI). For each block, the expression of the receive signal-to-interference-plus-noise ratio (SINR) of user $k$ is given by~:
\begin{equation}
\label{eq:SINR}
 \gamma_k = \frac{g_k p_k}{\displaystyle{\sigma^2+ \sum^K_{j\neq k} \theta_j g_j p_j}}
\end{equation}
where for all $j \in \mc{K}$, $g_j = |h_j|^2$ and $\theta_j$ represents a parameter depending on the interference scenario. For example, in a random code division multiple access (RCDMA) system with single-user decoding, we would have $\theta_j = \frac{1}{N}$ where $N$ is the spreading factor \cite{meshkati-jsac-2006}. This is the choice we will do. Nonetheless, note that the present work is not restricted to code division multiple access (CDMA) systems. Indeed, by choosing $\theta_j = 1$ (i.e., $N=1$) the above signal model corresponds to the information-theoretic channel model used for studying multiple access channels
\cite{wyner-it-1974,cover-book-1975}; in this setup, good channel codes are assumed (see e.g.,
\cite{belmega-twc-2009} for more comments on the multiple access
technique involved). Indeed, what matters the most in the model is that it captures the different aspects of the problem (especially the SINR structure). At last, the case where successive interference cancelation is used at the receiver ($\theta_j \in \{0,1\}$, depending on the decoding order) is left as an extension of the present work.

\subsection{Performance metric}

Assuming the above signal model, we assume that each transmitter wants to selfishly maximize the energy-efficiency of his communication with the receiver. The used performance metric is the one originally proposed in \cite{goodman-pc-2000}. For a given block, transmitter $k$ wants to maximize the following quantity~:
\begin{equation}
v_k(p_k,\mb{p}_{-k})=\frac{R_k \textsf{f}(\gamma_k)}{p_k} \ \text{[bit/J]}
\label{utility}
\end{equation}

where $R_k$ is the transmission rate (in bit/s), $\textsf{f}$ is an efficiency function (representing the block success rate), and the subscript $-k$ on vector $\mathbf{p}$ stands for ``all the transmitters except transmitter $k$'', i.e., $\mathbf{p}_{-k}=\left(p_1,\ldots,p_{k-1},p_{k+1},
\ldots,p_K\right)$. Note that, as a standard assumption, $R_k$ is assumed to be independent of $\gamma_k$ or $p_k$ which may correspond in practice to a given choice of modulation coding scheme. As motivated in \cite{Rodriguez-globecom-2003,belmega-TSP-2011}, the efficiency function is assumed to be an increasing and sigmoidal (or S-shaped) function verifying $0 \leq \textsf{f}(.) \leq 1$ with $\textsf{f}(0)=0$ and $\displaystyle{\lim_{x \rightarrow +\infty} \textsf{f}(x) = 1}$. The fact that $\textsf{f}$ is sigmoidal has at least two important consequences~: the utility function $v_k$ is quasi-concave w.r.t. $p_k$ and the derivative of $v_k$ vanishes at only one point which is different from $0$. We see that $R_k$ might be chosen to be SINR-dependent without affecting the problem analysis provided that the product $R_k \textsf{f}$ be a sigmoidal function. For the sake of clarity, we assume that the players have the same efficiency function $\textsf{f}$. In \cite{goodman-pc-2000} and related works, $p_k$ represents the power radiated by the transmitter. Interestingly, the above utility can also model situations where the power consumed by the whole transmitting device has to be accounted for. Indeed, by replacing the denominator of $u_k$ by $a p_k + b$, $(a,b)$ being a pair of non-negative constants, one obtains a first-order model of the device power consumption which includes both the consumption part which does not depend on the radiated power and the one due to the transmit power \cite{richter}. This does not change significantly the mathematical analysis of the power control problem. In order to focus our attention on the most important points of our analysis and make the exposition as clear as possible, the original model of \cite{goodman-pc-2000} has been selected (i.e., $(a,b)=(1,0)$).
\vspace{-0.5cm}
\subsection{Game-theoretic modeling~: review of the non-cooperative game of \cite{goodman-pc-2000}}
\label{sec:game}
An appropriate model for the power control problem described above is given by a strategic form game \cite{goodman-pc-2000}. A strategic form game consists of an ordered triplet comprising the set of players, their action or strategy sets, and their preference orders (or their utilities when they exist, which is the case here). The set of players is the set of transmitters $\mc{K}$, the action set is $\mc{P}_k = [0, P_k^{\max}]$, $k \in \mc{K}$, and the utility functions are defined by (\ref{utility}). This describes the model introduced by Goodman et al in \cite{goodman-pc-2000}. As \cite{goodman-pc-2000} and related references such as \cite{meshkati-jsac-2006}, the power levels are chosen to be continuous. This allows us to conduct a complete comparison analysis in terms of performance. However, this assumption is not always suited and the case of discrete power levels is therefore left as a complementary way of tackling the problems under investigation. An important solution concept for this game is the Nash equilibrium\footnote{App. \ref{sec:appGameTheory} reviews several game-theoretic notions. Note that the unconditional existence of a pure NE is, in part, a consequence of quasi-concavity for the utility functions.} (NE) \cite{Nash51}, which is a power profile/vector that is robust against unilateral deviations (no player has interest in deviating if the others keep the equilibrium strategy). The unique NE of this power control game is~:
\begin{equation}
p_k^{\mathtt{NE}}= \frac{\sigma^2}{g_k} \frac{\beta^\star}{1-\frac{K-1}{N}\beta^\star},\  k \in \mathcal{K}
\label{eq:NE-power}
\end{equation}
where $\beta^\star$ denotes the best SINR choice for user $k$ at the NE~; as explained in \cite{goodman-pc-2000}\cite{meshkati-jsac-2006}, a necessary condition for this equilibrium to be defined is that the system load is not too high ($\frac{K-1}{N} \beta^\star < 1$). Note that the equilibrium SINR is common to all users. It is easy to verify that $\beta^\star$ is the positive solution of the differential equation $x\textsf{f}'(x)-\textsf{f}(x)=0$, which is obtained by solving $\underset{x\neq0}{\max}\;\frac{\textsf{f}(x)}{x}$, i.e., an equivalent problem of $\underset{p_k}{\max}\; v_k$ (following from the assumption that $R_k$ is independent of $\gamma_k$ and $p_k$).

The equilibrium solution holds if the power constraint $p_k \leq P_k^{\max}$ is satisfied, which is what we will assume throughout this paper (see e.g., \cite{goodman-pc-2000} for further details about the case where the constraint is active). This game model, although leading to a decentralized solution in terms of decisions and CSI (see \cite{goodman-pc-2000}), has one main drawback~: the equilibrium solution can be inefficient. Interestingly, introducing some hierarchy between the players in terms of observation can improve the game outcome, as shown in \cite{lasaulce-twc-2009}. It turns out that hierarchy is naturally present in networks where some transmitters are equipped with a cognitive radio while the others are not. This is one of our motivations for formulating the problem in decentralized cognitive networks as a two-level Stackelberg game, with arbitrary numbers of cognitive radios, generalizing the $1$-leader
 $K-1$-follower game of \cite{goodman-pc-2000}. Compared to the latter reference \cite{goodman-pc-2000}, a second interesting feature of the game described below is that the cost induced by sensing is accounted for in the utility function of the cognitive transmitters. The proposed approach may be relevant in most applications where cognitive radio is useful. Indeed, one of the messages of this work is that if the fraction of transmitters who can observe their environment is too high, this may degrade the global performance. To mention an existing scenario where this type of approaches might be applied in the future, the case of WiFi systems can be mentioned. In France, operators provides more and more advanced access points. Typically, they want to optimize channel selection (which is a special case of power allocation) in a more and more efficient manner. Assuming that some access points (AP) are optimized according to a Nash strategies while others implement Stackelberg strategies allows one to provide a simplified model to account for the fact that advanced APs coexist with less advanced APs. Interestingly, as shown in this paper, as far as power control is concerned, having too many advanced APs might not be as good as the common sense would indicate.
\section{The two-level power control game with sensing costs}
\label{sec:power-control-game}
The set of transmitters $\mc{K}=\{1,2,...,K\}$ comprises $F$ terminals equipped with a cognitive radio while the $L=K-F$ other terminals have no sensing capabilities. The pair $(F,L)$ is assumed to be fixed throughout the whole section; it will be optimized in a centralized (resp. decentralized) manner in Sec. \ref{sec:centralized-l-f} (resp. Sec. \ref{sec:decentralized-l-f}). Without loss of generality, the set of non-cognitive (resp. cognitive) terminals will be $\mc{L}=\{1,2,...,L\}$ (resp. $\mc{F} = \{L+1, L+2, ..., K\}$). This two-level hierarchical game is played as follows. For each block, the non-cognitive transmitters (called the leaders) choose their power level rationally knowing that their decisions are going to be observed by the cognitive transmitters. The cognitive transmitters (called the followers) react to these decisions rationally. A choice is said to be rational in the sense that the transmitter maximizes his utility. To this end, we denote by $\mathbf{p}_{\mc{L}}\triangleq\left(p_1,\ldots,p_L\right)$ and $\mathbf{p}_{\mc{F}}\triangleq\left(p_{L+1},\ldots,p_K\right)$ the vectors of actions (transmit powers) of the leaders and followers, respectively. Also denote by $\mc U^{\star}(\mathbf{p}_{\mc{L}})$ the set of NE for the group of followers when the leaders play $\mathbf{p}_{\mc{L}}$. The resulting outcome of this interaction is a Stackelberg equilibrium (SE), which is defined as follows.

\begin{definition}[Stackelberg equilibrium] \emph{A vector of actions $\mathbf{p}^{\mathtt{SE}}=(\mb{p}^{\mt{SE}}_{\mc{L}},\mb{p}^{\mt{SE}}_{\mc{F}})$ is called a Stackelberg equilibrium, if $\mb{p}^{\mt{SE}}_{\mc{F}} \in \mc U^{\star}(\mb{p}^{\mt{SE}}_{\mc{L}})$ and the actions $\mb{p}^{\mt{SE}}_{\mc{L}}$ is a Nash equilibrium for the leaders\footnote{We assume, w.l.o.g. two players in a non-cooperative game with one leader and one follower. If the leader plays the NE action, then, as the follower observes this action and plays the best-response against it, the follower will play the NE strategy. Then, the NE strategy profile, if it exists, can be a SE. The behavior is the same when there are several leaders and followers. If the NE between the leaders corresponds to the NE of the game between leaders and followers, then the followers respond by playing the NE.
}.}
\end{definition}

By looking at the mathematical expression of the Stackelberg equilibrium defined above, we can see that if the NE exists, then SE also exists. But the SE is not included in the set of NE of a non-cooperative game. There exists several examples like the Cournot game in which the action chosen by the leader at the SE is different compared to the action chosen at the SE \cite{Fudenberg}. As the best-response of each player is a scalar-valued function (see \cite{lasaulce-twc-2009}), the determination of the Stackelberg equilibria of the game amounts to solving the following bi-level optimization problem:
\small
\begin{equation}
p_{\ell}^{\mathtt{SE}}\in \arg\max_{p_{\ell}} u_{\ell}\left(p_{\ell},\mb p_{-\ell}^{\mathtt{SE}},p_{L+1}^{\mathtt{SE}}(p_{\ell},\mb p_{-\ell}^{\mathtt{SE}}),\ldots,p_{K}^{\mathtt{SE}}(p_{\ell},\mb p_{-\ell}^{\mathtt{SE}})\right),\ \forall \ell \in \mc L
\end{equation}
\normalsize where for all $\mathbf{p}_{\mc{L}}$,
\small
\begin{eqnarray}
p_f^{\mathtt{SE}}(\mathbf{p}_{\mc{L}})&=&\arg\max_{p_f}u_\textsf{f}(\mathbf{p}_{\mc{L}},p_{L+1}^{\mathtt{SE}}
(\mathbf{p}_{\mc{L}}),\ldots,p_{f-1}^{\mathtt{SE}}(\mathbf{p}_{\mc{L}}),p_f,...,\nonumber \\
&&p_{f+1}^{\mathtt{SE}}(\mathbf{p}_{\mc{L}}),\ldots,p_{K}^{\mathtt{SE}}(\mathbf{p}_{\mc{L}})),\ \forall f \in \mc F
\end{eqnarray}
\normalsize
where the utility functions are given by~:
\begin{equation}
\label{eq:utility-def}
u_k(\bd{p}) =\left|
\begin{array}{ll}
v_k(\bd{p}) & \ \ \text{if } k \in \mc{L},\\
(1-\alpha_k) v_k(\bd{p}) & \ \ \text{if } k \in \mc{F}.
\end{array}
\right.
\end{equation}

The parameter $\alpha_k \in [0,1], \; k\in \mc{F}$ is a constant w.r.t. the power levels which accounts for the sensing cost (to be illustrative, we will choose $\alpha_k=\alpha$ in some places). This constant has no effect on the equilibrium strategies. However, when it will come to knowing whether being a follower or not, this constant will play a role. To elaborate further on this constant, it can be interpreted as the fraction of time a cognitive user $k\in\mc{F}$ spends for sensing. In order to have a good sensing capabilities\footnote{Under the assumption of single-user decoding (each useful signal is detected by considering the other signals as noise), a good sensing capability means that a follower can detect the existence of all the leaders. In particular, the leader whose link with the follower is the worst is detected.}, we assume that there exists a certain energy threshold $\xi_{\min}$ (see e.g., \cite{han-tvt-2004}) expressed in joule~:
\begin{equation*}
\label{eq:sensing_ineq}
\alpha_k T \min_{\ell\in\mc{L}}\left(g_{k\ell}p_{\ell}\right)\geq \xi_{\min}
\end{equation*}
where $T$ is the block duration in second and $p_{\ell}$ in Watt whereas $\alpha_k$ and $f(.)$ are unitless~; $g_{k\ell}$ is the channel gain between any leader $\ell\in\mc{L}$ and the considered follower $k\in \mc{F}$. If the above inequality holds, it means that the cognitive user $k\in \mc{F}$ is able to sense the presence of the primary ones. Apparently, we assume that the sensing constraint is feasible in the sense that there exists a minimum fraction $\alpha_k \leq 1,\; k\in \mc{F}$ above which the minimum energy threshold for sensing is attained. A necessary condition for this is that $T\; \ds{\min_{\ell\in\mc{L}}} \left(g_{k\ell}p_{\ell}\right) \geq \xi_{\min}$. For the sake of clarity, we suppose that the sensing cost is the same for every player $\alpha_k=\alpha,\;k\in \mc{F}$. At this point, the two-level hierarchical power control game is completely defined : the players are the $L$ non-cognitive transmitters and the $F$ cognitive transmitters, their action sets are $[0, P_k^{\max}]$, and their utilities are defined by (\ref{eq:utility-def}).

Following the standard methodology of equilibrium analysis (see e.g., \cite{lasaulce-spm-2009}\cite{lasaulce-book-2011}), three important issues to be dealt with are the existence, uniqueness, and efficiency issues for the Stackelberg equilibrium. The next theorem provides an element of response to the first two issues.

\begin{prop}[\cite{lasaulce-twc-2009}]
\label{prop:existence-uniq-se}
\textit{There always exists a Stackelberg equilibrium $\mb p^{\mathtt{SE}}$ in the two-level hierarchical game with $L\geq 1 $ leaders and $F\geq 1$ followers. The power profile defined by
\small
\begin{eqnarray*}
\forall \ell\in \mc L, \ \ p_{\ell}^{\mathtt{SE}}=\frac{\sigma^2}{g_{\ell}}\frac{N
\gamma_L^{\star}(N+\beta^{\star})}{N^2-N(F-1)\beta^{\star}
-\left[(N+\beta^{\star})(L-1)+F\beta^{\star}\right]\gamma^{\star}_L}\\
\ \forall f \in \mc F, \ \ p_f^{\mathtt{SE}}=\frac{\sigma^2}{g_f}
\frac{N\beta^{\star}(N+\gamma_L^{\star})}{N^2-N(F-1)\beta^{\star}
-\left[(N+\beta^{\star})(L-1)+F\beta^{\star}\right]\gamma^{\star}_L}
\end{eqnarray*}
\normalsize
is an SE and $\beta^{\star}$ is the positive root of $xf'(x)=\textsf{f}(x)$, and $\gamma^{\star}_L$ is the positive root of $x(1-\epsilon_L x)f'(x)=\textsf{f}(x)$, with
\begin{equation}
\epsilon_L=\frac{F\beta^{\star}}{N^2-N(F-1)\beta^{\star}}.\label{eq:epsilon_L}
\end{equation}
Moreover, the equilibrium $\mb p^{\mathtt{SE}}$ is unique if the following two conditions hold:
(1) $\underset{x\rightarrow 0^+}{\lim}\frac{f''(x)}{f'(x)} > 2\epsilon_L $, and (2) equation $x(1-\epsilon_Lx)f'(x)-\textsf{f}(x)=0$ has a single root in $(0,\beta^{\star})$.}
\end{prop}

Note that the existence of such an equilibrium is ensured from the properties of the Stackelberg game, especially the sigmoidness of $\textsf{f}$ \cite{lasaulce-twc-2009}. In \cite{lasaulce-twc-2009}, it also explained that the best-response of the players are scalar-valued functions, which facilitates the Stackelberg equilibrium analysis. Interestingly, the provided sufficient conditions for uniqueness can be checked to be satisfied for two typical efficiency functions used in the related literature namely, $\textsf{f}(x) = (1-e^x)^M$ and $\textsf{f}(x)= e^{-\frac{c}{x}}$, $c\geq0$  used in \cite{goodman-pc-2000} and \cite{belmega-TSP-2011} respectively.

At last but not least, we address the issue of efficiency for the derived Stackelberg equilibrium. The key point at stake is whether cognition helps to obtain a better decentralized network in terms of global energy-efficiency. For this purpose, we first compare the utility a player would get in a system where no cognitive transmitters exist with the one he would obtain in a system with cognitive transmitters (i.e., at the Nash equilibrium corresponding to \cite{goodman-pc-2000}). Our main results are summarized by the following proposition.

\begin{proposition}[SE versus NE]\label{prop:SEvsNE} \emph{The utility at the SE of the two-level hierarchical game with sensing cost of any leader is always greater or equal to the one obtained at the NE.}

\end{proposition}

If the cost for sensing is negligible, the next corollary follows.

\begin{corollary}[SE versus NE with no sensing cost]\label{coro:SEvsNE} \emph{The power profile at the SE Pareto dominates the power profile at the NE.}
\end{corollary}

The proof of Proposition \ref{prop:SEvsNE} and corollary \ref{coro:SEvsNE} are given in \cite{lasaulce-twc-2009}.
Another relevant question, initially raised in a context with a single leader and no sensing cost \cite{lasaulce-twc-2009} is whether it is better to follow or lead the power control game. Said otherwise, is it beneficial for a transmitter to be equipped with a cognitive radio when sensing costs are accounted for ? The answer is provided below.

\begin{proposition}[Following versus leading]\label{prop:FollowvsLead} \emph{At the Stackelberg equilibrium of the two-level power control game with sensing cost, a transmitter prefers to be a follower (that is to say, to sense) if the minimum energy threshold for sensing verifies~:}
\begin{equation}
\xi_{\min} \leq \left[\!1\!-\!\frac{\frac{\textsf{f}(\gamma_L^{\star})}{\gamma_L^{\star}}}{\frac{
\textsf{f}(\beta^{\star})}{\beta^{\star}}}\frac{(N\!
+\!\gamma_L^{\star})}{(N\!+\!\beta^{\star})}\!\right] \; T\;\min_{\ell  \in \mc{L}} g_f p_\ell^{\mathrm{SE}}.
\end{equation}
\end{proposition}

The proof of this result is given in App. \ref{sec:appProofPropCost}. Interestingly, it is possible to provide an explicit lower bound on the energy threshold for a cognitive radio for being energy-efficient. For a transmitter, this bound mathematically translates the tradeoff between the benefit (in terms of energy-efficiency) of being informed about the actions played by the others and the cost induced by acquiring this knowledge. If the sensing cost is negligible, then following becomes always better than leading, giving an incentive to equip a transmitter with a cognitive radio. However, if all the transmitters of the network are cognitive, the network energy-efficiency is not maximized, which is what is proved in the next section.

\vspace{-0.5cm}
\section{The sensing game}
\label{sec:sensing-game}

In the preceding section, the pair $(F,L)$ and the identities of cognitive and non-cognitive transmitters were fixed. A quite natural question is to ask whether the transmitters would effectively sense or not in a fully decentralized network where the decision to sense is left to them. Providing answers to this question is the purpose of this section. As a first step, we show the existence of an optimal number of cognitive transmitters in terms of social welfare (sum utility). The corresponding upper bound can be used to assess the price of anarchy of the network \cite{Papadimitriou-99} and therefore measuring the cost of having this decision decentralized. As a second step, we consider a sensing game in which each transmitter has two actions (sense/not sense). It is shown that each transmitter can learn his best decision provided that the number of blocks is sufficiently large.
\vspace{-0.5cm}
\subsection{On the optimal number of cognitive transmitters}
\label{sec:centralized-l-f}

The global energy-efficiency of the network is measured in terms of social welfare \cite{arrow-book-1963} or sum utility at the equilibrium which is defined by~:
\begin{equation}
w_L^{e} = \sum_{k=1}^K u_k = \sum_{k=1}^L u_k +  \sum_{k=L+1}^K u_k,
\label{swL}
\end{equation}
where $e \in \{\mathrm{SE}, \mathrm{NE}\}$. Note that a Nash equilibrium is obtained when $L=K$, indeed in this context, there is no followers and the game is no more hierarchical. The subscript $L$ has been added to the equilibrium profiles to clearly indicate that it is related to the number of leaders whenever $L \neq K$ (see equation (\ref{swL}). As the parameter $L$, which is the number of leaders or non-cognitive transmitters, belongs to a discrete set $\mc{K}$, the function $w_L : \mc{K} \rightarrow \mathbb{R}^+$ has necessarily a maximum\footnote{Note that, however, we do not try to optimize the identity of the followers or leaders with respect to their channel quality. This type of issues, which is of relevant in centralized scenarios, is addressed in \cite{lasaulce-twc-2009} in a special case.}. Is this maximum reached at the non-trivial points $L=1$ or $L=K$? >From Proposition \ref{prop:SEvsNE}, we know that it is not the case. Indeed, if the sensing cost is small enough, then the power profile at any SE Pareto-dominates the one of the NE obtained with $L=1$ or $L=K$. Thus, the latter points are not the maximizer candidates for the sum utility. However, when the sensing cost is arbitrary, answering the question analytically does not seem to be trivial. This is why we solve this maximization problem numerically in Sec. \ref{sec:numerical-results}. To still get some insights into the problem, we study a very special case of it. The interest in doing so is that it clearly shows that the sum utility maximizer is non-trivial and a little more about the connection with the network load can be learned.

We now consider the special case defined by the following four assumptions~:
\begin{itemize}
	\item Assumption 1~: $g_k=g, R_k=R$ for all $k \in \mathcal{K}$.
	\item Assumption 2~: $N>>(K-1)\beta^{\star}.$
    \item Assumption 3~: $\textsf{f}(x) = e^{-\frac{c}{x}}$, $c  \geq  0$.
    \item Assumption 4~: $1\leq L \leq K-1$.
\end{itemize}
The social welfare at the SE is given by~:
\begin{eqnarray}
w_L^{\mt{SE}}
&= & \sum_{\ell\in\mc{L}}  R_{\ell} \frac{\textsf{f}(\gamma_L^{\star})}{p^{\mt{SE}}_{\ell}} + \sum_{f\in\mc{F}} R_{f} \frac{\textsf{f}(\beta^{\star})}{p^{\mt{SE}}_f} \nonumber \\
&= & R \frac{g a_L}{\sigma^2}  \left[\frac{L\textsf{f}(\gamma_L^{\star})}{\gamma_L^{\star}(N+\beta^{\star})}+\frac{(K-L)\textsf{f}(\beta^{\star})}{\beta^{\star}(N+\gamma_L^{\star})}\right],
\label{eq:social_welfare}
\end{eqnarray}
where
\begin{equation}
a_L\triangleq N-(F-1)\beta^{\star}-\frac{\left[(N+\beta^{\star})(L-1)+F\beta^{\star}\right]
\gamma^{\star}_L}{N}.
\end{equation}
Note that in \cite{belmega-tsp-2010}, $c = 2^r -1$, where $r$ is the spectral efficiency of the used channel coding scheme (in bit/s per Hz). A possible choice is $r = \frac{R}{B}$ where $R$ is the data transmission rate and $B$ the required bandwidth to transmit. While Assumptions $2$ and $3$ are reasonable (they respectively correspond to a small system load and the case where the efficiency function is derived from the outage probability on the mutual information \cite{belmega-tsp-2010}), the first assumption is very strong and may happen in practice in specific scenarios e.g., in virtual multiple input multiple output (MIMO) networks with clusters of transmitters \cite{huh-tit-sub-2011} with similar flow types (a voice service typically). Indeed, if fast power control is considered namely, $g_k$ represents the fast fading, this symmetry assumption will almost never be verified in practice. Now, if slow power control is considered, $g_k$ may represent shadowing and path loss effects, and the $g_k'$s will be almost equal for all users being in a given neighborhood (as explained in \cite{huh-tit-sub-2011} where the notion of clusters of users is shown to be relevant). In any case, note that the symmetry assumption is very local and is only exploited to reinforce the existence of a non-trivial optimal number of followers/leaders and provide insights on this number. Again, the goal is not to claim a general expression for the optimal number of leaders. Rather, we just want to derive it in a special case to better understand the general optimization problem. Assumption $4$ is not restrictive since the cases $K=0$ and $K=L$ are ready. Under Assumption 2, we have that $a_L\approx N$. From  (\ref{eq:epsilon_L}) we have $\epsilon_L\approx 0$, which implies that $\gamma_L^{\star}\approx \beta^{\star}, \textsf{f}(\gamma_L^{\star})\approx \textsf{f}(\beta^{\star})$, and
\begin{equation}
\frac{\textsf{f}(\gamma_L^{\star})}{N+\beta^{\star}} \approx \frac{\textsf{f}(\beta^{\star})}{N+\gamma_L^{\star}}.
\end{equation}
This allows one to approximate the social welfare at the equilibrium (\ref{eq:social_welfare}) as~:
\begin{equation}
\widetilde{w}_L^{\mt{SE}}
\approx  R \frac{g \textsf{f}(\beta^{\star})N}{(N+\beta^{\star})\sigma^2} \left( \frac{L}{\gamma_L^{\star}}+\frac{K-L}{\beta^{\star}}\right).
\end{equation}

Note that the term $\frac{g\textsf{f}(\beta^{\star})N}{(N+\beta^{\star})\sigma^2}$ is independent of $L$, and the optimal solution $L^{\star}$ can be approximated by~:
\begin{equation}
\label{approx_maximization}
\tilde{L}^{\star} = \arg\max_L \left(\frac{L}{\gamma_L^{\star}}+\frac{K-L}{\beta^{\star}}\right).
\end{equation}
The main point in the above equation is that $\widetilde{L}^{\star}$ is related to $\gamma_L^{\star}$ which is the positive root of the equation $x(1-\epsilon_L x)\textsf{f}'(x)-\textsf{f}(x)=0$. It turns out that, under Assumption 3, a very simple expression for  $\gamma_L^{\star}$ can be obtained. One can easily check that~:
\begin{equation}
\gamma_L^{\star}=\frac{c}{1+\epsilon_L c}.
\end{equation}
Using the above, the optimization problem given by equation (\ref{approx_maximization}) boils down to~:
\begin{eqnarray*}
\widetilde{L}^{\star} &=& \arg \max_L \left( \frac{L}{\frac{c}{1+\epsilon_L c}}+\frac{K-L}{c} \right) = \arg \max_L \left(\epsilon_L L\right).
\end{eqnarray*}
Replacing the discrete variable $L$ with a non-negative real $\lambda$, the optimal solution of the corresponding concave function can be checked to be~:
\begin{equation}
\widetilde{\lambda}^{\star} =
\left|
\begin{array}{ccc}
\left(1+ \frac{N}{c} \right) \left(1 - \sqrt{1 - \frac{K}{N} \frac{c}{\frac{c}{N}+1}} \right) & \text{if} & \frac{K-1}{N} \leq \frac{1}{c}   \\
 K-  \kappa & \text{if} &  \frac{K-1}{N} > \frac{1}{c}
\end{array}\right.
\end{equation}
where $\kappa >0$ is arbitrary small (this constraint is added to meet the constraint $L\leq K-1$). Therefore, the optimal number of non-cognitive transmitters can be approximated  by $ \widetilde{L}^\star = \left\lfloor\widetilde{\lambda}^{\star}\right\rfloor$ or $\widetilde{L}^\star = \left\lceil \widetilde{\lambda}^{\star}\right\rceil$ depending on which number gives the maximal social welfare. From this expression of $\widetilde{L}^\star$, some interesting insights can be easily extracted. First of all, note that if the load is sufficiently small compared to $\frac{1}{\beta^\star}$ but at the same time greater than $\frac{1}{c}$, the social welfare is maximized for $\widetilde{L}^\star = K-1$. Indeed, when the load is small, the interaction between players is not strong and the impact of a hierarchy on the overall system is small (at least for a reasonably large system). Thus, the social welfare is maximized at the NE point, which corresponds to the case where all users are leader and play at the same time. A second type of insights can be obtained by considering the spectral efficiency related to the channel coding namely $c = 2^r - 1$. As already mentioned, note that it is assumed that the spectral efficiency involved by the multiple access technique is sufficiently small (Assumption 2). If we go further by assuming that both the multiple access technique spectral efficiency ($\frac{K}{N} \rightarrow 0$) and channel coding spectral efficiency ($c \rightarrow 0$) are small, we obtain that~:
\begin{equation}
\begin{array}{ccc}
\widetilde{\lambda}^\star & \approx &
\left(1+ \frac{N}{c} \right) \frac{1}{2} \frac{K}{N} \frac{c}{\frac{c}{N}+1} \approx  \frac{K}{2}
\end{array}
\end{equation}
which means that in the low spectral efficiency regime, the global energy-efficiency of the network is maximized when half of the transmitters are cognitive. If $c$ is large but the load is still small (meeting the constraint $\frac{K-1}{N} \leq \frac{1}{c}$) the same conclusion is obtained. Conducting a deep analysis to discuss the connections between spectral efficiency and energy-efficiency in decentralized cognitive networks in the general case (arbitrary load, with cost of sensing) seems to be a relevant and non-trivial extension of the simplified analysis provided here.

\subsection{Sensing game~: description and key property}
\label{sec:decentralized-l-f}

In the two-level hierarchical power control described in Sec. \ref{sec:power-control-game}, the
transmitter is, by construction, either a cognitive transmitter or a
non-cognitive one and the action of a player consists in choosing his transmit
power level. In fully decentralized networks, it is legitimate to ask about what a transmitter would decide between sensing (being cognitive or not) or not sensing. To analyze such a problem, we assume that two games occur sequentially for each block (this separation assumption will be fully justified in Sec. \ref{sec:EquilibriumAnalysis}). First, the transmitters decide to sense $(\textsf{S})$ or not to sense ($\textsf{NS})$. Then, their choose their power level based on their status (follower or leader) and therefore play the two-level power control game described in Sec.
\ref{sec:power-control-game}. The sensing game can therefore be described by a static game whose representation is given by the following triplet~:
\begin{eqnarray}
\mc{G}=\left(\mc{K},(\mathcal{A}_k)_{k\in \mc{K}},(U_k)_{k\in \mc{K}} \right)
\end{eqnarray}
where the actions sets are $\mathcal{A}_k = \mathcal{A}=\left\{\textsf{S}, \textsf{NS} \right\}$ and the utility functions are those obtained at the Stackelberg equilibrium of the power control game played in the second phase. If transmitter $k$ is a follower (i.e., he senses) and that there were $F$ followers during the sensing phase of the block then his utility is~:
\small
\begin{eqnarray*}
U_k(\textsf{S}, \bd{s}_{-k}^{(F,L)})
&=&   \frac{(1 - \alpha_k)g_k R_k}{\sigma^2}\frac{\textsf{f}(\beta^*)}{N\beta^{\star}(N+\gamma_{L+1}^{\star})} \times \\
&&\left\{N^2\!-\!N\beta^{\star} - \left[(N\!+\!\beta^{\star})L\!+ (F+1) \beta^{\star}\right]\gamma^{\star}_{L+1}\right\}
\end{eqnarray*}
\normalsize
where the notation $\bd{s}_{-k}^{(F,L)}$ means that there are $F-1$ followers and $L$ leaders in the set of players $\mc{K}\setminus{k}$. On the other hand, if transmitter $k$ chooses the action $\textsf{NS}$ he obtains~:
\small
 \begin{eqnarray*}
U_k(\textsf{NS}, \bd{s}_{-k}^{(F,L)})&=& \frac{g_k R_k}{\sigma^2}\frac{\textsf{f}(\gamma^*_L)}{N\gamma_{L+1}^{\star} (N + \beta^{\star})}\times \\
&&\left\{N^2\!-\!N\beta^{\star} - \left[(N\!+\!\beta^{\star})L\!+
(F+1) \beta^{\star}\right]\gamma^{\star}_{L+1}\right\}
\end{eqnarray*}
\normalsize
in terms of utility. The considered sensing game is a congestion game \cite{NisanRoughgardenTardosVazirani(AlgoGameTheory)07} (and therefore a potential game) under strong conditions but is always a weighted potential game. The latter property is known to be very useful for studying existence of pure NE and convergence of learning algorithms or distributed iterative algorithms towards NE. For instance, in \cite{Monderer-Shapley-1996,Monderer-Shapley-1996b}, Monderer and Shapley proved that every weighted potential game has the Fictitious Play\footnote{The FP learning algorithm can be found in the quoted references or \cite{young-book-2004} but essentially it consists in assuming that each player observes the actions of the others and maximizes his average utility based on the empirical frequencies of use of actions of the others.} Property (FPP). This guarantee that every learning algorithm that is Fictitious Play process converges in belief to equilibrium.
All of this is the purpose of the remaining of this section. For making this paper sufficiently self-containing we review several useful definitions concerning potential games \cite{Monderer-Shapley-1996}.

\begin{definition}[Monderer and Shapley 1996 \cite{Monderer-Shapley-1996}]
\emph{The strategic form game $\mc{G}$ is a potential game if there is a
potential function $V : \mathcal{A} \longrightarrow \mathbb{R}$ such that}
\begin{eqnarray*}
U_k(s_k,\bs{s}_{-k}) -  U_k(s_k',\bs{s}_{-k}) = V(s_k,\mb{s}_{-k}) -  V(s_k',\bs{s}_{-k}),\\
\quad \forall k\in \mc{K},\; s_k,s_k'\in \mathcal{A}_k.
\end{eqnarray*}
\end{definition}

\begin{theorem}\label{theo:PotentialGame}
\emph{The sensing game $\mc{G}=(\mc{K},(\mathcal{A})_{k\in \mc{K}},(U_k)_{k\in \mc{K}})$ is an
exact potential game if and only if one of the two following
conditions is satisfied~:}
\small
\begin{eqnarray}
\forall i,j\in \mc{K}\quad R_ig_i&=&R_jg_j,\label{eq:ExactPotential1}\\
U_L(F+1,L+1) - U_L(F,L+2)&=&(1-\alpha)(U_F (F+2,L)-\nonumber\\
&&U_F(F+1,L+1)), \label{eq:ExactPotential2}
\end{eqnarray}
\normalsize
\emph{where $U_L(F+1,L+1)$ is defined by $\frac{\sigma^2U_i(F+1,L+1)}{R_ig_i}$ when player $i\in \mc{K}$ is one of the $F+1$ followers and
$U_F(F+1,L+1)$ is defined by $\frac{\sigma^2 U_i(F+1,L+1)}{R_ig_i}$ when player $i\in \mc{K}$ one of the $L+1$ leaders.}
\end{theorem}

Condition (\ref{eq:ExactPotential1}) is a (strong) symmetry condition and is obtained under Assumption 1 (Sec. \ref{sec:centralized-l-f}), which would be reasonable for a cluster of transmitters in a virtual MIMO network with a common service (e.g., voice). In fact, it is more realistic not to make these assumptions and claim for a potential property which is sufficient for key issues such as convergence of some important learning dynamics.

\begin{definition}[Monderer and Shapley 1996 \cite{Monderer-Shapley-1996}]
\emph{The strategic form game $\mc{G}$ is a weighted potential game if there is a
vector $(\mu_i)_{i\in \mc{K}}$ and a potential function $V : \mathcal{A}
\longrightarrow \mathbb{R}$ such that~:}
\small
\begin{eqnarray*}
\forall i\in \mc{K},\; (s_i,s_i')\in \mathcal{A}_i^2,&& \nonumber\\
 U_i(s_i,s_{-i}) -  U_i(s_i',s_{-i}) &=& \mu_i(V(s_i,s_{-i}) -  V(s_i',s_{-i})).
\end{eqnarray*}
\normalsize
\end{definition}

It turns out that such a vector can be found.

\begin{theorem}\label{theo:WeightedPotential}
\emph{The sensing game $\mc{G}=(\mc{K},(\mathcal{A}_i)_{i\in \mc{K}},(U_i)_{i\in \mc{K}})$ is
a weighted potential game with the weight vector~:}
\begin{eqnarray}
\forall i\in \mc{K}, \quad \mu_i = \frac{R_ig_i}{\sigma^2}.
\end{eqnarray}
\end{theorem}
The proof is given in App. \ref{sec:appProofTheoWeigh}.
\vspace{-0.5cm}
\subsection{Equilibrium analysis}
\label{sec:EquilibriumAnalysis}

\subsubsection{Existence}

First of all, note that since the sensing game is finite (i.e., both the
number of players and the sets of actions are finite), the existence
of at least one mixed NE is guaranteed
\cite{nash-academy-50}. Now, since the game is a weighted
potential game, the existence of at least one pure NE is guaranteed
\cite{Monderer-Shapley-1996}. We might restrict our attention to pure and mixed Nash equilibria. However, as it will be clearly seen in the 2-player case study (Sec. \ref{sec:Efficiency}), this may pose a problem of fairness and efficiency. This is the main reason
why we also study the set of correlated equilibria (App. \ref{sec:appGameTheoryCE}) of the sensing game. The concept of correlated equilibrium \cite{Aum74}
allows one to enlarge the set of equilibrium utilities. Every utility vector inside the convex hull of the Nash equilibrium utilities is a correlated equilibrium, which guarantees the existence of correlated equilibria in general.

\subsubsection{Uniqueness}

Here, we provide a brief analysis of uniqueness for the pure NE. This matters since pure NE are attractors of important dynamics such as the replicator dynamics (which corresponds to the limit of important learning schemes) \cite{borkar08}. One obvious advantage of having uniqueness of the game outcome is to make the game predictable, which may be useful from a designer standpoint. As mentioned above, by contrast, the number of correlated equilibria is generally greater than one and more typically infinite. The following proposition provides sufficient condition under which the sensing game (always with costs) has a unique pure NE.

\begin{proposition}
\emph{Assume the following two conditions are satisfied~:}
\small
\begin{eqnarray}
\alpha &>& 1 - \frac{ (N+\gamma_{{K-1}}^{\star})\left(N - (K - 1)\beta^{\star}\right)}{ N^2\!-\!N\beta^{\star} - \left[(N\!+\!\beta^{\star}){(K-1)}\!+ 2 \beta^{\star}\right]\gamma^{\star}_{{K-1}}}\label{eq:CondUniqu1}\\
 \alpha&>& 1 - \frac{\gamma_{K-1}^{\star} (N + \beta^{\star})\textsf{f}(\beta^{\star})}{\beta^{\star}(N+\gamma_{K-1}^{\star})\textsf{f}(\gamma^{\star}_{K-1} )}\label{eq:CondUniqu2}.
\end{eqnarray}
\normalsize
Then the unique Nash equilibrium of the game is $(s_1^*,s_2^*,...,s_K^*) =
 (\textsf{NS}, \textsf{NS},..., \textsf{NS})$.
\end{proposition}

Condition (\ref{eq:CondUniqu1}) insures that the non-sensing strategy $\textsf{NS}$ dominates the sensing strategy $\textsf{S}$ when none of the other player sense. Condition (\ref{eq:CondUniqu2}) insures that the non-sensing strategy $\textsf{NS}$ dominates the sensing strategy $\textsf{S}$ when some of the other player sense. Both conditions together imply that the sensing strategy $\textsf{S}$ is always a dominated strategy for each player. The unique Nash equilibrium of the game is $(s_1^*,s_2^*,...,s_K^*) =  (\textsf{NS}, \textsf{NS},..., \textsf{NS})$.



\subsubsection{Efficiency}\label{sec:Efficiency}

In a decentralized network, since no or little coordination between terminals is available, an important issue is the efficiency of the network at the equilibrium state. Are the mixed or pure NE of the sensing game efficient in terms of utility? To be illustrative and to understand in a deep manner the problem under investigation, our choice, in this section, is to mainly focus on the $2-$transmitter case but most of the provided results can be extended to the general case $K\geq 2$.


\begin{theorem}\label{theo:NumberNash}[Number of NE] \emph{The matrix game has the following NE~:}
\begin{itemize}
  \item \emph{a unique NE if and only if}
$
(C1): \alpha > \frac{\beta^* - \gamma^*}{1-\beta^*\gamma^*};
$

  \item \emph{three NE if and only if}
$
(C2): \alpha<\frac{\beta^* - \gamma^*}{1-\beta^*\gamma^*};
$

  \item \emph{an infinite number of NE if}
$
(C3): \alpha = \frac{\beta^* - \gamma^*}{1-\beta^*\gamma^*}.
$
\end{itemize}
\end{theorem}

The proof of this result is provided in App. \ref{sec:appNumberNashEqu}. There is also a strictly mixed equilibrium which can be found using the indifference principle. Let $(x,1-x)$ the mixed strategy for player 1 and $(y,1-y)$ the mixed strategy for player 2. As proven in App. \ref{sec:appStrictlyMixedEqu}, there is a unique pair $(x^*,y^*)$ satisfying the indifference principle. The corresponding distribution is given by~:
\small
\begin{eqnarray}\label{eq:mixedEquilibrium}
x^*=y^* = \frac{(1-\alpha)\frac{\textsf{f}(\beta^*)}{\beta^*}(1-\beta^*) - \frac{\textsf{f}(\gamma^*)}{\gamma^*} \frac{1- \gamma^*\beta^*}{ 1+ \beta^*}  }{X},
\end{eqnarray}
\normalsize
with $X=(1-\alpha)\frac{\textsf{f}(\beta^*)}{\beta^*}(1-\beta^*) - \frac{\textsf{f}(\gamma^*)}{\gamma^*} \frac{1- \gamma^*\beta^*}{ 1+ \beta^*}   + \frac{\textsf{f}(\beta^*)}{\beta^*}(1-\beta^*) - (1-\alpha) \frac{\textsf{f}(\beta^*)}{\beta^*} \frac{1- \gamma^*\beta^*}{ 1+ \gamma^*}$ and the corresponding equilibrium utilities are~:
\begin{eqnarray*}
U_1(x^*,y^*) &=& \frac{R_1g_1}{\sigma^2} \upsilon\\
U_2(x^*,y^*) &=& \frac{R_2g_2}{\sigma^2} \upsilon
\end{eqnarray*}
with
\begin{eqnarray*}
  \upsilon &=& \frac{(1-\alpha)\frac{\textsf{f}(\beta^*)}{\beta^*}(1-\beta^*)\frac{\textsf{f}(\beta^*)}{\beta^*}(1-\beta^*) 
}{X}-\\
&&\frac{\frac{\textsf{f}(\gamma^*)}{\gamma^*} \frac{1- \gamma^*\beta^*}{ 1+ \beta^*} (1-\alpha) \frac{\textsf{f}(\beta^*)}{\beta^*} \frac{1- \gamma^*\beta^*}{ 1+ \gamma^*}}{X}.\\
\end{eqnarray*}

\begin{figure}
	\centering
		 \includegraphics[width=0.3\textwidth]{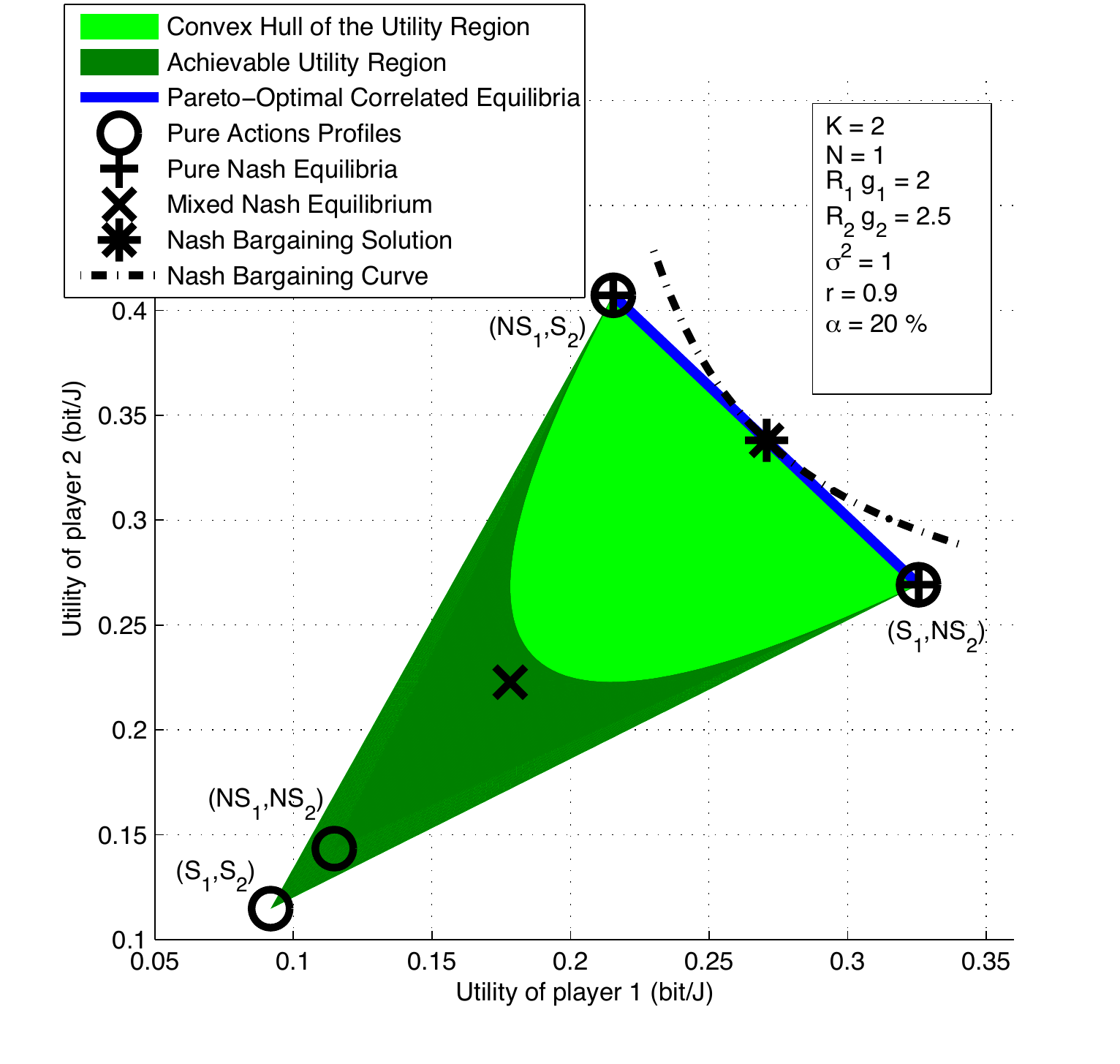}
	\caption{The figure depicts, for a $2$-transmitter scenario, the region of achievable utilities of the sensing game where each transmitter decides whether to sense of not to sense. The figure also shows the different equilibrium points of the game. One of the messages of this figure is the interest in terms of fairness in stimulating a correlated equilibrium instead of a Nash equilibrium. In particular a Nash bargaining solution can be obtained.}	 \label{fig:2playerMatrix}
\end{figure}

Fig. \ref{fig:2playerMatrix} represents the three equilibrium utility points for a typical scenario. The shaded area represents the region of feasible utilities for a given scenario (described under Fig. \ref{fig:2playerMatrix}). Operating at one of the pure NE can be unfair for one of the transmitters and therefore inefficient for a certain fairness criterion \cite{kelly98}. Operating at the mixed NE is clearly suboptimal since it is Pareto-dominated by some feasible pairs of utilities. A way of dealing with fairness or/and Pareto-inefficiencies is to induce correlated equilibria (CE) in the game.

In practice, having a correlated equilibrium means that the players have no interest in ignoring (public or private) signals which would recommend them to play according to certain joint distribution over the action profiles of the game. In wireless networks, a correlated equilibrium can be induced by a common signalling
from a source which is exogenous to the game. It may be a signal generated by the receiver itself but also an FM (frequency modulation) signal, or a GPS (global positioning system) signal, meaning that the additional cost for adding this signal may be zero if the terminals are already able to decode such a signal. At last, note that such a coordination mechanism is scalable in the sense that it can accommodate a high number of transmitters; in practice, physical limitations may arise e.g., if the signal is sampled into a finite number of bits. If $\alpha > \frac{\beta^* - \gamma^*}{1-\beta^*\gamma^*}$, as there is only one NE, the convex hull of NE boils down to a point and there does not exist any other correlated equilibrium other than this NE. Rather, we assume that the sensing cost verifies condition (C1)
which is the case of interest since several NE exist (see Theorem \ref{theo:NumberNash}). In this case the following result holds.

\begin{theorem}
\emph{Any convex combination of NE is a CE. In particular, if there exists a utility vector $\nu=(\nu_1,\nu_2)$  and a parameter $\lambda\in [0,1]$ such that~:}
\begin{eqnarray}
\nu_1 &=&  \lambda U_1(\textsf{S}_1,\textsf{NS}_2) + (1 - \lambda ) U_1(\textsf{NS}_1,\textsf{S}_2)\\
\nu_2 &=&  \lambda U_2(\textsf{S}_1,\textsf{NS}_2) + (1 - \lambda ) U_2(\textsf{NS}_1,\textsf{S}_2),
\end{eqnarray}
then $\nu$ is a correlated equilibrium.
\end{theorem}
Clearly, a signal recommending the transmitters to play the action profile $(\textsf{S}_1,\textsf{NS}_2)$ (resp. $(\textsf{NS}_1,\textsf{S}_2)$) for a fraction of the time equals to $\lambda$ (resp. to $1-\lambda$) induces a correlated equilibrium. This specific signalling structure leads to the set of equilibria represented by the bold segment in Fig. \ref{fig:2playerMatrix}. The figure illustrates the potential gains which can be obtained by implementing a simple coordination mechanism in the sensing game with costs.

We would like to end this section dedicated to the efficiency of the equilibria of the game by mentioning the potential sub-optimality induced by playing the sensing game and power control game separately (in two consecutive phases). Indeed, it would be legitimate to ask about what would happen if a transmitter were deciding jointly whether to sense or not and his power level. In such a case the action set of a transmitter would be~:
\begin{eqnarray}
\widetilde{\mc{A}_k}=  \{\textsf{S}_k, \textsf{NS}_k\}\times [0, P_k^{\max}].
\end{eqnarray}
An action $a_k = (s_k, p_k)$ has therefore two components. The first component is discrete whereas the second component is continuous. This framework is referred to an hybrid control in control theory \cite{EvansSavkin99,LiberzonMorse99}. While the control theory literature is rich concerning hybrid control, this is not the case for hybrid control games. In particular, general existence theorems for Nash equilibria seem to be unavailable. This is one of the reasons we will only consider the special case of two transmitters. In the $2-$player hybrid control game it can be easily seen that the two pure NE of the sensing game are no longer equilibria in this new game. Instead, we have the following result.

\begin{proposition}
\emph{The unique Nash equilibrium of the $2-$player hybrid control game is given by~:}
\begin{equation}
(a_1^*, a_2^*) = (\textsf{NS}_1, p_1^{\mathrm{NE}}, \textsf{NS}_2, p_1^{\mathrm{NE}})
\end{equation}
\emph{where $p_k^{\mathrm{NE}}$ is given by (\ref{eq:NE-power}).}
\end{proposition}

This result immediately follows from the fact that action every action under the form $(\textsf{S}, p_k)$ is dominated by the action $(\textsf{NS}, p_k)$. Although the proof of this result is trivial, the interpretation is nonetheless interesting. It shows the existence of a Braess paradox in the hybrid control game~: although the players have more options in the hybrid game, the equilibrium utilities are less than those obtained in the separated case where they first decide to sense or not and then adapt their power level. In additional to implementation considerations, this gives us another reason to perform the decision process in two consecutive phases.

\section{Numerical results}
 \label{sec:numerical-results}

\begin{figure}
	\centering
		 \includegraphics[width=0.3\textwidth]{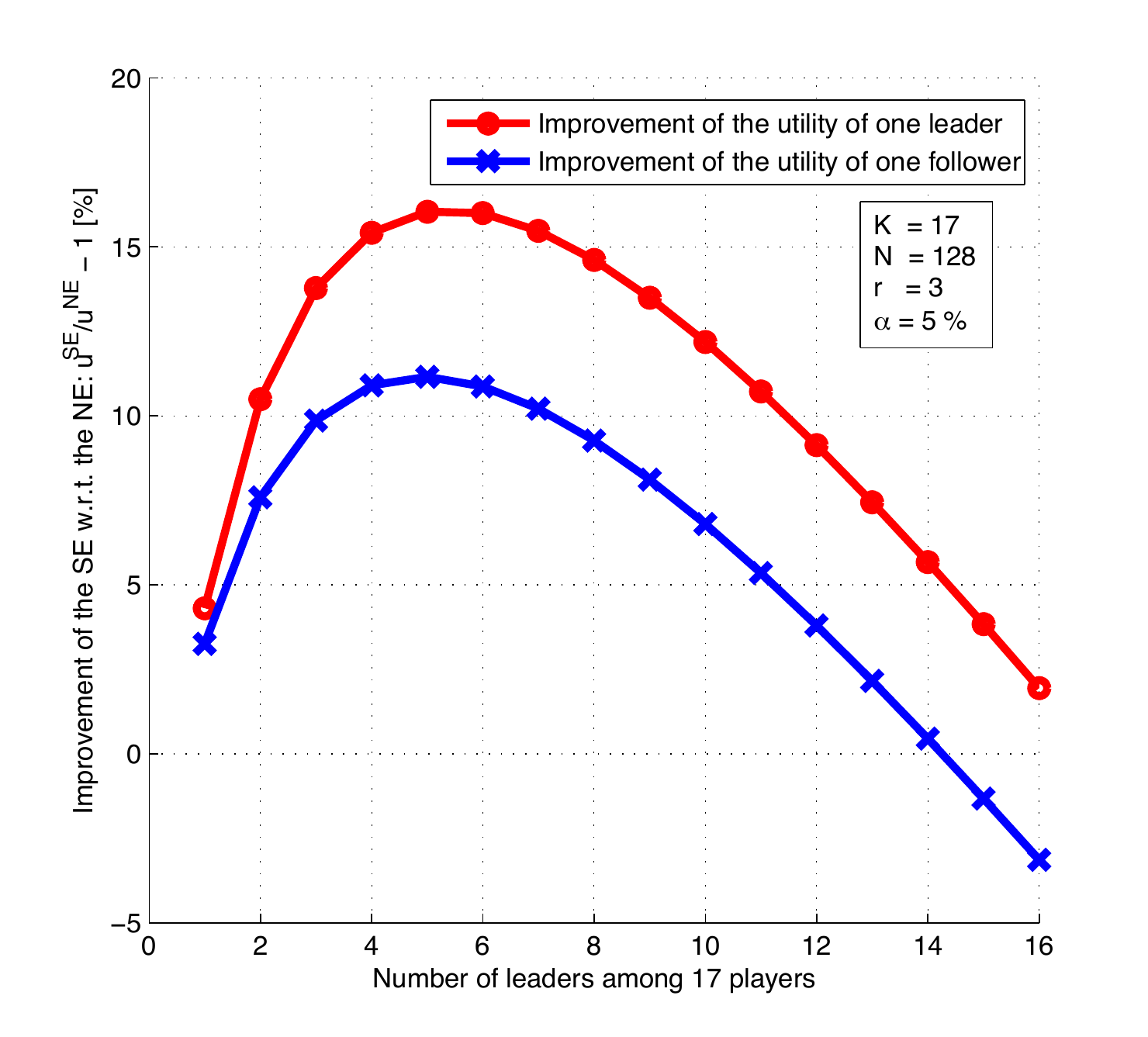}
	\caption{This figure represents the relative gain (in $\%$) in terms of individual energy-efficiency obtained by equipping $F=K-L$ transmitters with a cognitive radio. For typical scenarios, we see that maximizing the number of cognitive transmitters is not optimal. On the other hand, if there is only one cognitive radio ($F=1$ or $L=16$, as assumed in \cite{lasaulce-twc-2009}) one can degrade the individual performance for a typical value for the sensing cost ($5\%$ of the time-slot is spent for sensing).} 	 \label{fig:UtilitiesFL}
\end{figure}

\begin{figure}
	\centering
		 \includegraphics[width=0.3\textwidth]{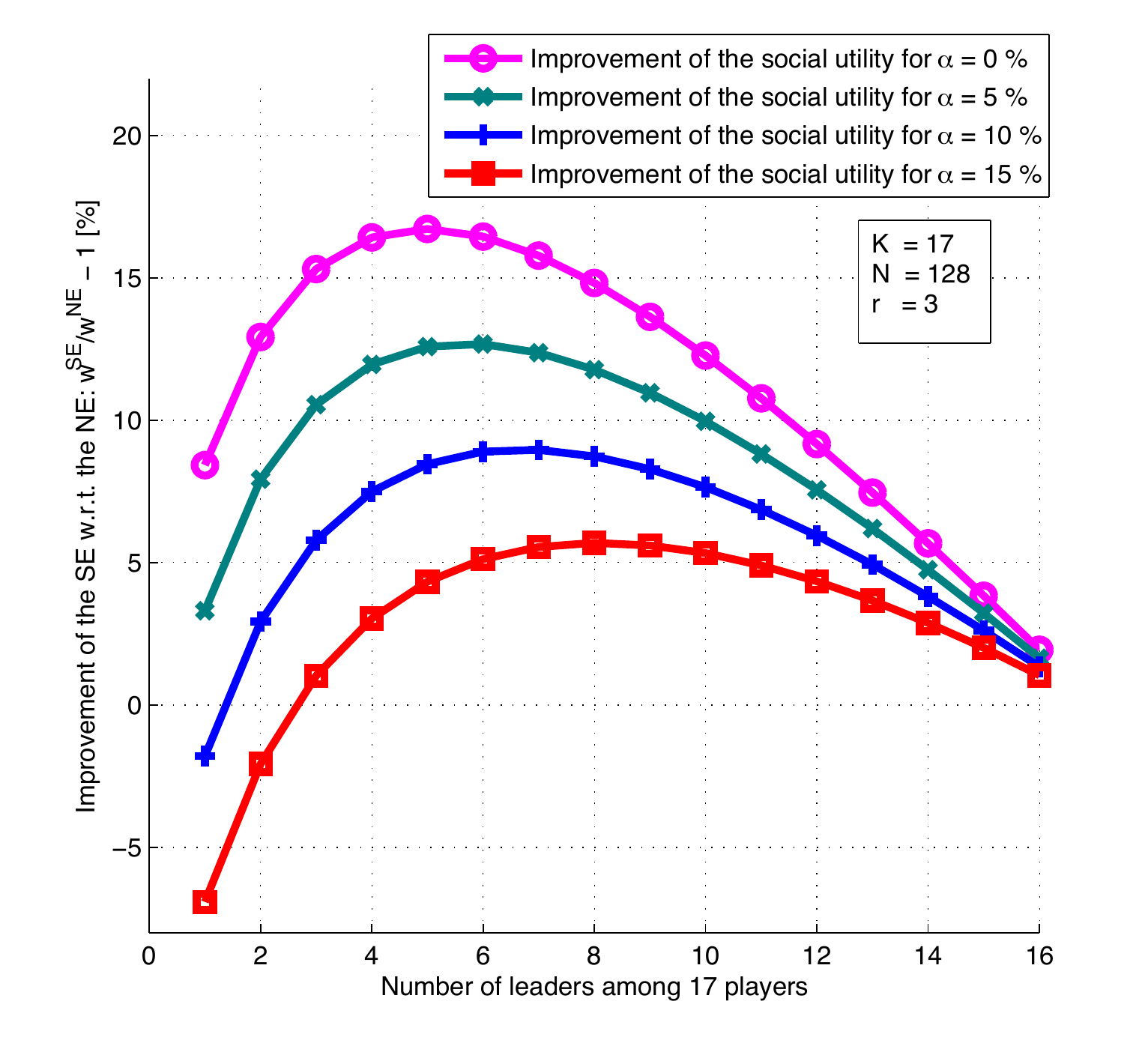}
		\caption{This figure represents the improvement in terms of utility sum or network energy-efficiency at the Stackelberg equilibrium compared to the case of Nash equilibrium (i.e., no transmitter is equipped cognitive radio) with $K=17$ players, $N=128$, the efficiency function is $f(x)=e^{-\frac{2^r - 1}{x}}$ with $r = 3$ bit/s per Hz and for different numbers of leaders. The different curves corresponds to different values for the sensing cost : $ \alpha\in \{0\%,5\%,10\%,15\%\} $. When $5\%$ of the time has to be spent for sensing, the network energy-efficiency can be improved by $13\%$ whereas it is $17\%$ when this cost is close to zero.} \label{fig:SumUtilities}
\end{figure}

\begin{figure}
	\centering
\includegraphics[width=0.3\textwidth]{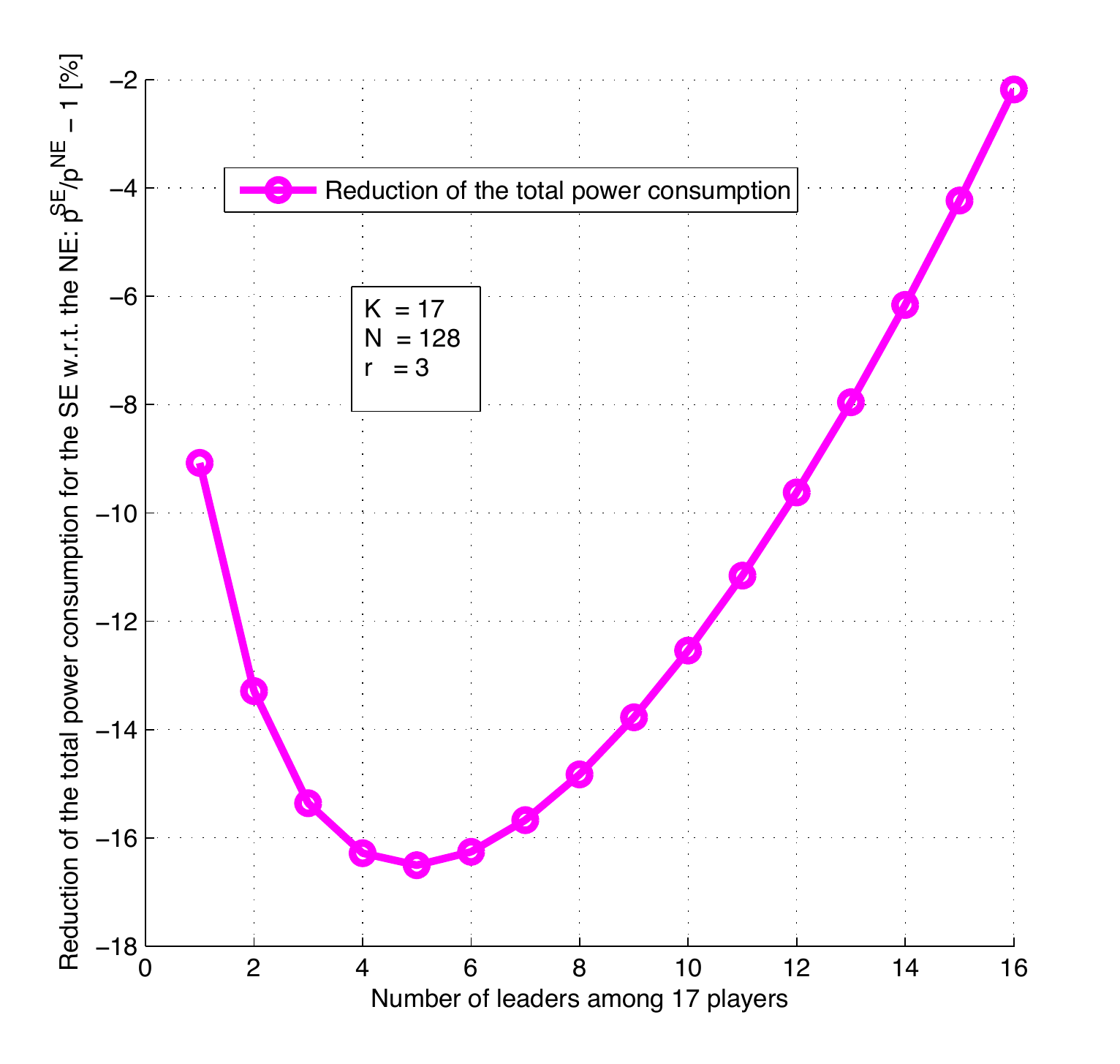}
	\caption{From previous figures, we know that equipping the network with an adequate number of cognitive radios can significantly improve the network energy-efficiency. It turns out that not only efficiency is maximized by doing so but that the total consumed power is also reduced. Scenario illustrated~: $K=17$ players, $N=128$, the efficiency function is $f(x)=e^{-\frac{2^r - 1}{x}}$ with $r = 3$ for different numbers of leaders.}	 \label{fig:Power}
\end{figure}

\begin{figure}
	\centering
		 \includegraphics[width=0.3\textwidth]{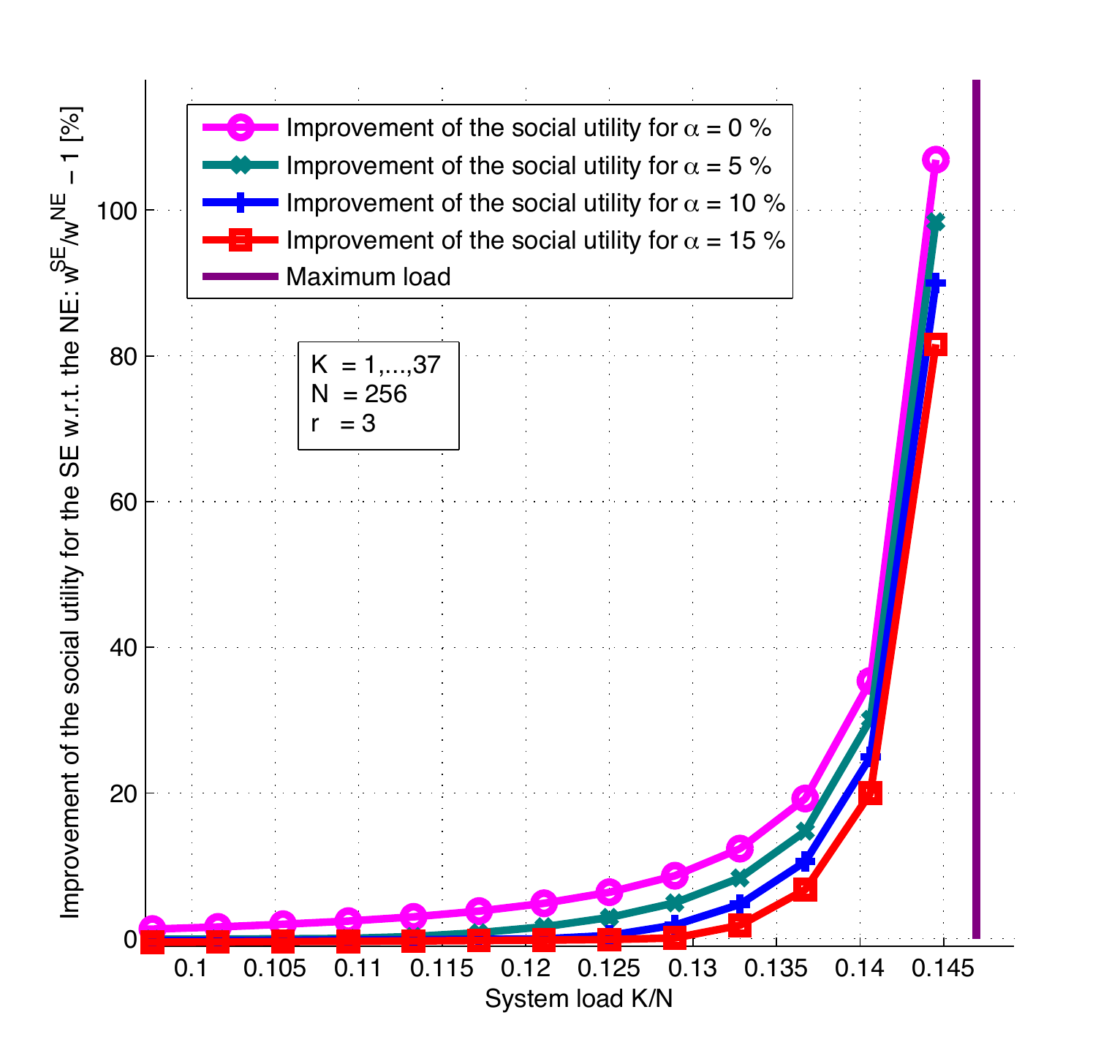}
	\caption{The objective of this figure is to show the relationship between the maximum gain obtained in terms of network energy-efficiency and the spectral efficiency in terms of user load $(K/N)$ (the spectral efficiency in terms of channel coding is always fixed and set to $r=3$ bit/s per Hz). From this figure, it is seen that there is a strong interest in operating at a load level close to the maximum limit tolerated by the system (ensuring the existence of an equilibrium, see Sec. \ref{sec:game}).}\label{fig:Load}
\end{figure}

In this section, numerical results are provided to validate our theoretical claims. Note that, although simple scenarios considered, the authors believe that most of messages and insights conveyed by the present numerical analysis hold in more advanced simulation setups e.g., considering standardized channel modulation and coding schemes (MCS), real frequency selective channel impulse responses, imperfect channel state information, and sensing techniques accounting for estimation noise. Indeed, as explained in \cite{goodman-pc-2000}, the choice of a specific MCS will generally lead to a packet success rate having the assumed properties. As shown in \cite{meshkati-jsac-2006}, the case of frequency selective channels is treatable once the frequency flat case has been treated. Therefore, only the impact of channel estimation noise seems to be more uncertain and would call for a more challenging extension of the results provided here. We consider a random CDMA scenario with a spreading factor equal to $N$ and the efficiency function is chosen to be $f(x)=e^{-\frac{2^r - 1}{x}}$ with different parameters $r$ \cite{belmega-TSP-2011}. We consider two scenarios. The first scenario is provided in Fig. \ref{fig:2playerMatrix}. This scenario provides a clear understanding of the variety of equilibria in the sensing game. The pure, mixed, and correlated equilibria are represented on the utility region.
The utility region of the sensing game with two players $K=2$, no spreading $N=1$, the sensing cost $ \alpha=20\% $, the sigmoidal function  $f(x)=e^{-\frac{2^r - 1}{x}}$ with $r = 0.9$ and the following parameters~: $ R_1\;g_1 = 2,\;R_2\;g_2 = 2.5,\;\sigma^2=1$. The pure actions lead to the utilities marked by circles, the dark green region corresponds to the pair of utilities that is achievable with mixed actions whereas the light green region corresponds to the utilities that are achievable only with correlated actions. The two pure equilibrium utilities, denoted by $+$, correspond to both upper left extremal pure utilities, the completely mixed equilibrium utility, denoted by $\times$, is located in the interior of the dark green region. The blue line between the two pure equilibria represents a sub-set of correlated equilibrium utilities that corresponds to the Pareto-optimal frontier. The Nash bargaining solution, denoted by $*$, corresponds to the intersection of the hyperbolic curve with the set of correlated equilibrium utilities. It provides a fair and optimal equilibrium solution for the sensing game.\\
The second scenario considers a sensing game with 17 players, $N=128$ sub-carriers, the sensing cost $ \alpha=5\% $, the sigmoidal function  $f(x)=e^{-\frac{2^r - 1}{x}}$. For simplicity, we assume an homogeneous scenario in terms of transmission rate $R_k=R$ and $r = \frac{R}{B} 3$ bit/s per Hz for different numbers of leaders. Note that the value for $R$ will not matter since only normalized/relative performance gains will be considered. The seemingly non-typical choice for $K$ results from typical choices on the other parameters. Indeed, when fixing the spreading factor to $N=128$ (typical e.g., in cellular systems), the spectral efficiency to $r=3$ bit/s per Hz (typical in cellular systems as well), one finds that the maximum number of admissible users for the Nash equilibrium to be implemented is $18$ (see the denominator of (4))~: $r=3 \Rightarrow c= 7  \Rightarrow \beta^* = 7 \Rightarrow
    \frac{K-1}{N} < \frac{1}{7} \Rightarrow K < 18$ where $f(x) = e^{-\frac{c}{x}}$, $c=2^r - 1$, $\beta^*$ is the unique solution of $x f'(x) - f(x) = 0$. Fig. \ref{fig:UtilitiesFL} and \ref{fig:SumUtilities} allows one to evaluate the improvement brought by the Stackelberg approach compared to the Nash equilibrium approach for the utility of one leader, one follower and the social utility. The sensing cost influences the results in two ways. First, the sensing cost affects the gain obtained by the follower compared to the leader. Indeed, in this figure, the improvement of a leader is always larger than the improvement of one follower, which is not true in general. Second, the improvement of the utility of one follower compared to the Nash equilibrium utility is negative when the number of leader is strictly more than 14. In that case, the sensing cost is compensated by the improvement due to the Stackelberg approach compared to the Nash equilibrium approach. The optimal number of leaders is 5 when considering both the improvement of the utility of one leader or the improvement of the utility of one follower. The improvement of the sum utility of the Stackelberg equilibrium compared to the social utility at the Nash equilibrium for the sensing game is given in figure \ref{fig:SumUtilities}. The sensing cost decreases the improvement of the social utility. This is especially the case when the number of follower is large and the number of leader is small. The Stackelberg approach provides up to $16.5\%$ of improvement compared to the Nash equilibrium approach. \\
Fig. \ref{fig:Power} illustrates how much the total power consumption can be reduced for the sensing game with $K=17$ players, $N=128$ sub-carriers, the sigmoidal function  $f(x)=e^{-\frac{2^r - 1}{x}}$ with $r = 3$ for different numbers of leaders~: note that at the same time, the energy-efficiency is optimized. The best reduction of the power consumption is achieved with the number of leaders is 5, the reduction of the power consumption is more than $16\%$. We observe that the total power reduction is maximum with the number of cognitive users maximizes the social welfare of the system.
Finally, the figure \ref{fig:Load} represents the improvement of the maximal social welfare (depending on the number of cognitive users) of the sensing game compared to the social welfare at the Nash equilibrium solution, depending on the load $(K/N)$ of the system. The four curves correspond to different sensing cost $ \alpha\in \{0\%,5\%,10\%,15\%\} $. When the load approaches its maximal value $1/\beta^{\star}+1/N$, the improvement of the social utility is greater than $100\%$. Then, we can conclude that our hierarchical framework with optimal number of cognitive equipments becomes more efficient in terms of social utility when a cognitive wireless network is high loaded.

\section{Conclusion}
\label{Conclusion}

%

In this paper, we have introduced a new power control game where the
action of a player is hybrid, one component of the action is discrete while the
other is continuous. The first component is discrete since it corresponds to deciding whether to sense the radio environment or not~; the second component is continuous because it corresponds to choosing the transmit power level in an interval. Whereas the general study of hybrid games is of independent and game-theoretic interest and remains to be done, it turns out that in our case we can prove the
existence of a kind of Braess paradox which allows us to restrict our
attention to two separate games played consecutively~: choosing the discrete and continuous actions jointly is less efficient than choosing them separately over time. The power control game is studied in detail and it is shown that there exists an optimal number of cognitive transmitters which maximizes the network utility, meaning that introducing too much cognition is not globally energy-efficient. This holds whether the cost of sensing is set to zero or not. From an individual point of view, the intuition which consists in saying that sensing is beneficial only if the sensing cost is acceptable, can be proved. As distributed networks are considered, global efficiency of the network is generally not guaranteed. Equilibria are indeed less energy-efficient (say in terms of sum utility) than the centralized solution. The (hierarchical) approach we propose can therefore be seen as a tradeoff in terms of global performance and required signaling. Conducting a refined analysis in terms of signaling for the power control problem would be relevant. On the other hand, the sensing game can be shown to have desirable properties like being weighted potential. This is a key property since many learning algorithms are known to converge in such games, proving that this decision can be learned over time with partial information only. Additionally, this game is shown to have a non-trivial set of correlated equilibria. These equilibria are very useful since they allow one to introduce some fairness among the transmitters and can be stimulated by a public signal incurring no cost in terms of extra signalling from the receiver~; in this respect the famous Nash bargaining solution (used in the wireless literature for having both a fair and cooperative solution, see e.g., \cite{larsson-jsac-2008}\cite{ma-jcta-2011}) can be reached. This work therefore provides several new results of practical interest for cognitive wireless networks but, of course, the proposed concepts would need to be developed further to make them more appealing in terms of implementability. In particular, technical issues related to spectrum usage might be considered by introducing frequency selective channels and the corresponding power allocation problem. Considering a more general structure of interference networks, the relevance of successive interference cancelation in terms of energy-efficiency might be assessed. Of course, classical issues such as the impact of channel estimation is also of practical interest, especially regarding to the fact that some learning algorithms are known to be robust against this type of errors.

\appendices

\section{Review of some game-theoretic concepts}\label{sec:appGameTheory}
%
%

\subsection{Correlated Equilibria}\label{sec:appGameTheoryCE}
In this section, we provide the definition of correlated equilibrium (CE).
This equilibrium concept was introduced by Aumann in  \cite{Aum87} and extends the concept of Nash equilibrium.
Correlated equilibria are used in section \ref{sec:Efficiency} in order to provide more fair equilibrium solutions.
\begin{definition}
\emph{A probability distribution $Q\in \Delta(\mc{A})$ is a canonical
correlated equilibrium if for each player $i$, for each action
$a_i\in \mc{A}_i$ that satisfies $Q(a_i)>0$ we have~:}
\small
\begin{eqnarray}
\sum_{a_{-i} \in \mc{A}_{-i}} Q(a_{-i} \mid a_i)u_i(a_i,a_{-i})\geq \nonumber\\
\sum_{a{-i} \in \mc{A}_{-i}}Q(a_{-i} \mid a_i)u_i(b_i,a_{-i}),\quad \forall b_i\in \mc{A}_i.\label{CorrelatedLinearInequlity}
\end{eqnarray}
\normalsize
\end{definition}
The result of Aumann 1987 \cite{Aum87} states that for any correlated
equilibrium, it corresponds to a canonical correlated equilibrium.
\begin{theorem}[Aumann 1987, prop. 2.3 \cite{Aum87}]\label{theorem:correlated}
The utility vector $u$ is a correlated equilibrium utility if and
only if there exists a distribution $Q\in \Delta(\mc{A})$ satisfying the
linear inequality contraint (\ref{CorrelatedLinearInequlity}) with $u=
E_Q U$.
\end{theorem}

\vspace{-0.5cm}
\subsection{Potential of the Sensing Game}\label{sec:appGameTheoryPotential}
In this section, we provide the potential function of the sensing game presented in section \ref{sec:sensing-game}.
\begin{theorem}
\emph{The equilibria of the above potential game are the maximizers
of the Rosenthal potential function \cite{Rosenthal73}.}
\begin{eqnarray*}
\small
&&\{S = (\mc{A}_1,\ldots,\mc{A}_K) | S\in NE\}
= \arg\max_{(F,L)} \Phi(F,L)\\
 &=& \arg\max_{(F,L)}\left[ (1-\alpha)\sum_{i=1}^F U(S,i,K-i) \right.\\
 &&\left.+ \sum_{j=1}^L U(NS,K-j,j) \right]
 \normalsize
\end{eqnarray*}
\end{theorem}
The proof follows directly from the one of Rosenthal's theorem \cite{Rosenthal73}. Let us simplify the expression of the potential function, which gives:
\small
\begin{eqnarray*}
\Phi(F,L) &=& \left[ (1-\alpha)\sum_{i=1}^F U(S,i,K-i) + \sum_{j=1}^L U(NS,K-j,j) \right]\\
&=&  (1-\alpha)\sum_{i=1}^F \frac{g_iR_i}{\sigma^2}\frac{\textsf{f}(\beta^{\star})}{N\beta^{\star}(N+\gamma_{(K-i)}^{\star})}\left(N^2\!-\!N\beta^{\star}-\right.\\
&&\left.\left[(N\!+\!\beta^{\star})(K-i)\!+ (i+1) \beta^{\star}\right]\gamma^{\star}_{(K-i)}\right)\\
&+& \sum_{j=1}^L \frac{g_iR_i}{\sigma^2}\frac{\textsf{f}(\gamma^{\star}_j)}{N\gamma_{j}^{\star} (N + \beta^{\star})}\left(N^2\!-\!N\beta^{\star} - \right.\\
&&\left.\left[(N\!+\!\beta^{\star})j\!+ (K-j+1) \beta^{\star}\right]\gamma^{\star}_{j}\right).
\end{eqnarray*}
\normalsize

\section{Proof of proposition \ref{prop:FollowvsLead}}\label{sec:appProofPropCost}

In this section, we prove the proposition \ref{prop:FollowvsLead}.
In order to have good sensing capabilities, there exists a certain energy threshold $\xi_{\min}$ such that:
\begin{eqnarray*}
\alpha\; T \;\min_{\ell\in\mc{L}}\left(g_{f\ell}p_{\ell}\right)\geq \xi_{\min}
\Longleftrightarrow \alpha \geq \frac{\xi_{\min}}{T \;\min_{\ell\in\mc{L}}\left(g_{f\ell}p_{\ell}\right)}
\end{eqnarray*}
where $T$ is the block duration, $g_{f\ell}$ is the channel gain between any leader $\ell\in \mc{L}$ and the considered follower $f\in \mc{F}$, $p_{\ell}$ is the power level of the leader $\ell\in \mc{L}$ and $\alpha$ is the sensing cost. The utility of the follower is maximized when the sensing cost $\alpha$ is minimal, that is $\alpha^{\star}=\frac{\xi_{\min}}{T \min_{\ell\in\mc{L}}\left(g_{f\ell}p_{\ell}\right)}$.

\begin{eqnarray*}
&&\frac{\max_{\alpha\in [0,1]}U_f^{\mathtt{SE}}(\alpha)}{U_l^{\mathtt{SE}}}\geq1\\
&\Longleftrightarrow& \frac{(1-\alpha^{\star})\frac{R_k\textsf{f}(\beta^{\star})}{p_f^{\mathtt{SE}}}}{\frac{R_k\textsf{f}(\beta^{\star})}{p_l^{\mathtt{SE}}}}\geq 1 \\
&\Longleftrightarrow& \bigg(1-\frac{\xi_{\min}}{T \;\min_{\ell\in\mc{L}}\left(g_{f\ell}p_{\ell}\right)}\bigg)\frac{\frac{\textsf{f}(\beta^{\star})}{\beta^{\star}}  (N+\beta^{\star})}
{\frac{\textsf{f}(\gamma_L^{\star})}{\gamma_L^{\star}}(N+\gamma_L^{\star})}\geq1\\
&\Longleftrightarrow& \frac{\frac{\textsf{f}(\beta^{\star})}{\beta^{\star}}  (N+\beta^{\star})}
{\frac{\textsf{f}(\gamma_L^{\star})}{\gamma_L^{\star}}(N+\gamma_L^{\star})}- 1 \geq\frac{\xi_{\min}}{T \;\min_{\ell\in\mc{L}}\left(g_{f\ell}p_{\ell}\right)} \frac{\frac{\textsf{f}(\beta^{\star})}{\beta^{\star}}  (N+\beta^{\star})}
{\frac{\textsf{f}(\gamma_L^{\star})}{\gamma_L^{\star}}(N+\gamma_L^{\star})}\\
&\Longleftrightarrow&\bigg(1 -  \frac{\frac{\textsf{f}(\gamma_L^{\star})}{\gamma_L^{\star}}(N+\gamma_L^{\star})}{\frac{\textsf{f}(\beta^{\star})}{\beta^{\star}}  (N+\beta^{\star})} \bigg)\geq\frac{\xi_{\min}}{T \;\min_{\ell\in\mc{L}}\left(g_{f\ell}p_{\ell}\right)} \\
&\Longleftrightarrow&T \;\min_{\ell\in\mc{L}}\left(g_{f\ell}p_{\ell}\right) \bigg(1 -  \frac{\frac{\textsf{f}(\gamma_L^{\star})}{\gamma_L^{\star}}(N+\gamma_L^{\star})}{\frac{\textsf{f}(\beta^{\star})}{\beta^{\star}}  (N+\beta^{\star})} \bigg)\geq\xi_{\min}
\end{eqnarray*}
This concludes the proof of Proposition \ref{prop:FollowvsLead}.

\section{Proof of Theorem \ref{theo:NumberNash}}\label{sec:appNumberNashEqu}
In this section, we characterize the equilibria of the $2-$player sensing
game. The first important remark is that the Nash equilibrium utilities
are always dominated by the Stackelberg equilibrium utilities. This
implies that the following equation holds for any parameters $\alpha\geq0$.
\begin{eqnarray*}
U_1(\textsf{NS}_1,\textsf{S}_2) \leq U_1(\textsf{S}_1,\textsf{S}_2)\\
U_2(\textsf{S}_1,\textsf{NS}_2) \leq U_2(\textsf{S}_1,\textsf{S}_2)
\end{eqnarray*}
Thus the action $(\textsf{S}_1,\textsf{S}_2)$ is not an equilibrium of the game.
To compute the equilibria of this game, it remains to compute the following differences:
\begin{eqnarray*}
U_1(\textsf{NS}_1,\textsf{NS}_2) - U_1(\textsf{S}_1,\textsf{NS}_2)\\
U_2(\textsf{NS}_1,\textsf{NS}_2) - U_2(\textsf{NS}_1,\textsf{S}_2)
\end{eqnarray*}
The above differences are equal and does not depend on a particular player.
We provide the proof of Theorem \ref{theo:NumberNash}.
Suppose first that condition (C2) is met.
\begin{eqnarray*}
&&\alpha<\frac{\beta^{\star} - \gamma^{\star}}{1-\beta^{\star}\gamma^{\star}}\\
&\Longleftrightarrow & \frac{1-\gamma^{\star}\beta^{\star} - \beta^{\star}-\gamma^{\star} }{(1-\gamma^{\star}\beta^{\star})} <  1- \alpha \\
&\Longleftrightarrow & \frac{\textsf{f}(\beta^{\star})}{\beta^{\star}}(1-\beta^{\star}) < (1 - \alpha)\frac{\textsf{f}(\beta^{\star})}{\beta^{\star}}\frac{ 1 - \beta^{\star}\gamma^{\star}}{ 1 +\gamma^{\star}} \\
&\Longleftrightarrow & \frac{R_ig_i\textsf{f}(\beta^{\star})(1-\beta^{\star})}{\sigma^2\beta^{\star}} < (1-\alpha)\frac{R_ig_i\textsf{f}(\beta^{\star})(1-\gamma^{\star}\beta^{\star})}{\sigma^2\beta^{\star}(1+\gamma^{\star})}
\end{eqnarray*}
The last inequality implies that the games has two pures equilibria  $(\textsf{NS}_1,\textsf{S}_2)$, $(\textsf{S}_1,\textsf{NS}_2)$ and one strictly mixed equilibrium $(x^*,y^*)$ defined by the equations (\ref{eq:mixedEquilibrium}). If condition (C1) is satisfied, then the strategies $(\textsf{S}_1)$ and $(\textsf{S}_2)$ are dominated and then the game has one pure equilibrium $(\textsf{NS}_1,\textsf{NS}_2)$ and if condition (C3) is met, the game has an infinite number of NE.

\section{Mixed Nash Equilibria}\label{sec:appStrictlyMixedEqu}

If condition (C2) is met, the sensing
game has a strictly mixed equilibrium.
In this section, we provide a characterization of the mixed equilibrium strategies $(x^*,y^*)$ using the indifference principle.

\small
\begin{eqnarray*}
&&\frac{R_1g_1\textsf{f}(\beta^{\star})(1-\beta^{\star})}{\sigma^2\beta^{\star}} \cdot y_2 +  \frac{R_1g_1\textsf{f}(\gamma^{\star})(1-\gamma^{\star}\beta^{\star})}{\sigma^2\gamma^{\star}(1+\beta^{\star})} \cdot (1- y_2) \\
&=&  (1-\alpha)\frac{R_1g_1\textsf{f}(\beta^{\star})(1-\gamma^{\star}\beta^{\star})}{\sigma^2\beta^{\star}(1+\gamma^{\star})}  \cdot y_2 +\\
&& (1-\alpha)\frac{R_1g_1\textsf{f}(\beta^{\star})(1-\beta^{\star})}{\sigma^2\beta^{\star}}  \cdot (1- y_2), 
\end{eqnarray*}
which is equivalent to:
\begin{eqnarray*}
&&y_2  \cdot [ \frac{R_1g_1\textsf{f}(\beta^{\star})(1-\beta^{\star})}{\sigma^2\beta^{\star}}  -  (1-\alpha)\frac{R_1g_1\textsf{f}(\beta^{\star})(1-\gamma^{\star}\beta^{\star})}{\sigma^2\beta^{\star}(1+\gamma^{\star})}\\
&&+ (1-\alpha)\frac{R_1g_1\textsf{f}(\beta^{\star})(1-\beta^{\star})}{\sigma^2\beta^{\star}} -  \frac{R_1g_1\textsf{f}(\gamma^{\star})(1-\gamma^{\star}\beta^{\star})}{\sigma^2\gamma^{\star}(1+\beta^{\star})} ] \\
&=&
(1-\alpha)\frac{R_1g_1\textsf{f}(\beta^{\star})(1-\beta^{\star})}{\sigma^2\beta^{\star}}
- \frac{R_1g_1\textsf{f}(\gamma^{\star})(1-\gamma^{\star}\beta^{\star})}{\sigma^2\gamma^{\star}(1+\beta^{\star})}.
\end{eqnarray*}
Then we obtain:
\begin{eqnarray*}
y_2 &=& \frac{(1-\alpha)\frac{\textsf{f}(\beta^{\star})}{\beta^{\star}}(1-\beta^{\star}) - \frac{\textsf{f}(\gamma^{\star})}{\gamma^{\star}} \frac{1- \gamma^{\star}\beta^{\star}}{ 1+ \beta^{\star}}  }{ X },
\end{eqnarray*}
\normalsize
with $X=(1-\alpha)\frac{\textsf{f}(\beta^{\star})}{\beta^{\star}}(1-\beta^{\star}) - \frac{\textsf{f}(\gamma^{\star})}{\gamma^{\star}} \frac{1- \gamma^{\star}\beta^{\star}}{ 1+ \beta^{\star}}   + \frac{\textsf{f}(\beta^{\star})}{\beta^{\star}}(1-\beta^{\star}) - (1-\alpha) \frac{\textsf{f}(\beta^{\star})}{\beta^{\star}} \frac{1- \gamma^{\star}\beta^{\star}}{ 1+ \gamma^{\star}}$.
Replacing the above $y_2$ into the indifference equation, we obtain the utility of player 1 at the mixed equilibrium.
\begin{eqnarray*}
U_1(x_2,y_2) 
&=& \frac{R_1g_1}{\sigma^2} \left(\frac{(1-\alpha)\frac{\textsf{f}(\beta^{\star})}{\beta^{\star}}(1-\beta^{\star})\frac{\textsf{f}(\beta^{\star})}{\beta^{\star}}(1-\beta^{\star})}{X}\right.\\
&& \left. -\frac{ \frac{\textsf{f}(\gamma^{\star})}{\gamma^{\star}} \frac{1- \gamma^{\star}\beta^{\star}}{ 1+ \beta^{\star}} (1-\alpha) \frac{\textsf{f}(\beta^{\star})}{\beta^{\star}} \frac{1- \gamma^{\star}\beta^{\star}}{ 1+ \gamma^{\star}}}
{X}\right),
\end{eqnarray*}
with $X=(1-\alpha)\frac{\textsf{f}(\beta^{\star})}{\beta^{\star}}(1-\beta^{\star}) - \frac{\textsf{f}(\gamma^{\star})}{\gamma^{\star}} \frac{1- \gamma^{\star}\beta^{\star}}{ 1+ \beta^{\star}}   + \frac{\textsf{f}(\beta^{\star})}{\beta^{\star}}(1-\beta^{\star}) - (1-\alpha) \frac{\textsf{f}(\beta^{\star})}{\beta^{\star}} \frac{1- \gamma^*\beta^{\star}}{ 1+ \gamma^{\star}} $.
The same argument applies:
\begin{eqnarray*}
U_2(x_2,y_2) &=& \frac{R_2g_2}{\sigma^2} \left(\frac{(1-\alpha)\frac{\textsf{f}(\beta^{\star})}{\beta^{\star}}(1-\beta^{\star})\frac{\textsf{f}(\beta^{\star})}{\beta^{\star}}(1-\beta^{\star})}{X}\right.\\
&& - \left.\frac{\frac{\textsf{f}(\gamma^{\star})}{\gamma^{\star}} \frac{1- \gamma^{\star}\beta^{\star}}{ 1+ \beta2} (1-\alpha) \frac{\textsf{f}(\beta^{\star})}{\beta^{\star}} \frac{1- \gamma^{\star}\beta^{\star}}{ 1+ \gamma^{\star}}
}{X}\right).
\end{eqnarray*}


%
%
%

\section{Proof of Theorem \ref{theo:PotentialGame}}\label{sec:appProofTheoPotential}

The proof of Theorem \ref{theo:PotentialGame} uses the corollary 2.9 of the article \cite{Monderer-Shapley-1996} (see also the theorem 3.1 in \cite{Sandholm10}).
\begin{theorem}
\emph{The game $G$ is a potential game if and only if for every player $i,j\in \mc{K}$, every pair of actions $s_i,t_i\in \mc{A}_i$ and $s_j,t_j\in \mc{A}_j$ and every joint action $s_k\in \mc{A}_{K\backslash \{i,j\}}$, we have that}
\small
\begin{eqnarray*}
&& U_i(t_i,s_j,s_k)- U_i(s_i,s_j,s_k) + U_i(s_i,t_j,s_k) -U_i(t_i,t_j,s_k) +\\
&& U_j(t_i,t_j,s_k)- U_j(t_i,s_j,s_k) +U_j(s_i,s_j,s_k) -U_j(s_i,t_j,s_k) =0
\end{eqnarray*}
\end{theorem}
\normalsize
Let us prove that the  two conditions provided by our theorem are equivalent to the one of corollary 2.9 in \cite{Monderer-Shapley-1996}. We introduce the following notation defined for each player $i\in \mc{K}$ and each action $T\in \mathcal{A}$.
$$
\mu_i=\frac{R_ig_i}{\sigma^2}\quad \mbox{and} \quad U^T(t_i,t_j,s_k)=\frac{U_i^T(t_i,t_j,s_k)}{\mu_i}.
$$
For every player $i,j\in \mc{K}$, every pair of actions $s_i,t_i\in \mc{A}_i$ and $s_j,t_j\in \mc{A}_j$ and every joint action $s_k\in S_{K\backslash \{i,j\}}$, we have the following equivalences:
\scriptsize
\begin{eqnarray*}
U_i(\textsf{NS}_i,\textsf{S}_j,\textsf{S}_k)- U_i(\textsf{S}_i,\textsf{S}_j,\textsf{S}_k)+ U_i(\textsf{S}_i,\textsf{NS}_j,\textsf{S}_k) -U_i(\textsf{NS}_i,\textsf{NS}_j,\textsf{S}_k)=\\
U_j(\textsf{NS}_i,\textsf{S}_j,\textsf{S}_k)- U_j(\textsf{NS}_i,\textsf{NS}_j,\textsf{S}_k)+U_j(\textsf{S}_i,\textsf{NS}_j,\textsf{S}_k) - U_j(\textsf{S}_i,\textsf{S}_j,\textsf{S}_k)\\
\Longleftrightarrow
\begin{cases}
\mu_i = \mu_j \\
U_L(F+1,L+1) - U_L(F,L+2) \\
+   (1-\alpha)(U_F(F+1,L+1) - U_F (F+2,L))=0
\end{cases}
\end{eqnarray*}
\normalsize
Thus the sensing game is a potential game if and only if one of the two following conditions is satisfied~:
\begin{eqnarray*}
\small
\forall (i,j)\in \mc{K},\quad R_ig_i&=&R_jg_j,\\
U_L(F+1,L+1) - U_L(F,L+2)&=&\\
(1-\alpha)[U_F (F+2,L)&-&U_F(F+1,L+1)].
\end{eqnarray*}
\normalsize
\section{Proof of Theorem \ref{theo:WeightedPotential}}\label{sec:appProofTheoWeigh}
The proof of this theorem \ref{theo:WeightedPotential} follows the same line of the proof in App. \ref{sec:appProofTheoPotential}. It suffices to show that the auxiliary game defined as follows is a potential game.
\begin{eqnarray}
\widetilde{G}=(K,(\mathcal{A})_{i\in \mc{K}},(\tilde{U}_i)_{i\in \mc{K}})
\end{eqnarray}
Where the utility are defined by the following equations with $\mu_i = \frac{R_ig_i}{\sigma^2}$.
\begin{eqnarray}
\tilde{U}_i(s_i,s_{-i})&=&\frac{U_i(s_i,s_{-i})}{\mu_i}
\end{eqnarray}
From the above demonstration, it is easy to show that, for every player $i,j\in \mc{K}$, every pair of actions $s_i,t_i\in \mc{A}_i$ and $s_j,t_j\in \mc{A}_j$ and every joint action $s_k\in \mc{A}_{K\backslash \{i,j\}}$, we have the following equality:
\small
\begin{eqnarray*}
&&\tilde{U}_i(t_i,s_j,s_k)- \tilde{U}_i(s_i,s_j,s_k)+\tilde{U}_i(s_i,t_j,s_k) -\tilde{U}_i(t_i,t_j,s_k)=\\
&& \tilde{U}_j(t_i,s_j,s_k)-\tilde{U}_j(t_i,t_j,s_k)+ \tilde{U}_j(s_i,t_j,s_k)-\tilde{U}_j(s_i,s_j,s_k).
\end{eqnarray*}
\normalsize
Using Corollary 2.9 in \cite{Monderer-Shapley-1996}, we conclude that the sensing game is a weighted potential game.




\begin{IEEEbiography}[{\includegraphics[width=1in,height=1.25in,clip,keepaspectratio]{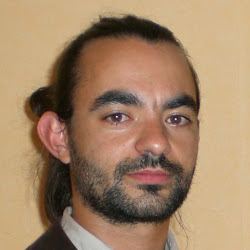}}]{Mael Le Treust}
Mael Le Treust earned his Dipl\^ome dÕEtude Approfondies (M.Sc.) degree in Optimization, Game Theory and Economics (OJME) from the Universit\' e de Paris VI (UPMC), France in 2008 and his Ph.D. degree from the Universit\' e de Paris Sud XI in 2011, at the Laboratoire des signaux et systmes (joint laboratory of CNRS, Sup\' elec, Universit\' e de Paris Sud XI) in Gif-sur-Yvette, France. He was a post-doctoral researcher at the Institut d'\' electronique et d'informatique Gaspard Monge (Universit\' e Paris-Est) in Marne-la-Vall\' ee, France and he is currently a post-doctoral researcher at the Centre Energie, Mat\' eriaux et T\' el\' ecommunication (Universit\' e INRS) in Montr\' eal, Canada. He was also a Math T.A. at the Universit\' e de Paris I (Panth\' eon-Sorbonne) and Universit\' e de Paris VI (UPMC), France. His research interests are information theory, game theory and wireless communications.
\end{IEEEbiography}

\begin{IEEEbiography}[{\includegraphics[width=1in,height=1.25in,clip,keepaspectratio]{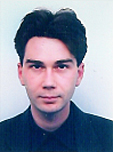}}]
{Samson Lasaulce}
Prof. Samson Lasaulce received his BSc and AgrŽgation degree in Physics from ƒcole Normale SupŽrieure (Cachan) and his MSc and PhD in Signal Processing from ƒcole Nationale SupŽrieure des TŽlŽcommunications (Paris). He has been working with Motorola Labs for three years (1999, 2000, 2001) and with France TŽlŽcom R\&D for two years (2002, 2003). Since 2004, he has joined the CNRS and SupŽlec as a Senior Researcher. Since 2004, he is also Professor at ƒcole Polytechnique. His broad interests lie in the areas of communications, information theory, signal processing with a special emphasis on game theory for communications. Samson Lasaulce is the recipient of the 2007 ACM VALUETOOLS, 2009 IEEE CROWNCOM, 2012 ACM VALUETOOLS best student paper award and the 2011 IEEE NETGCOOP best paper award. He is an author of the book ?Game Theory and Learning for Wireless Networks: Fundamentals and Applications?. He is currently an Associate Editor of the IEEE Transactions on Signal Processing. 
\end{IEEEbiography}

\begin{IEEEbiography}[{\includegraphics[width=1in,height=1.25in,clip,keepaspectratio]{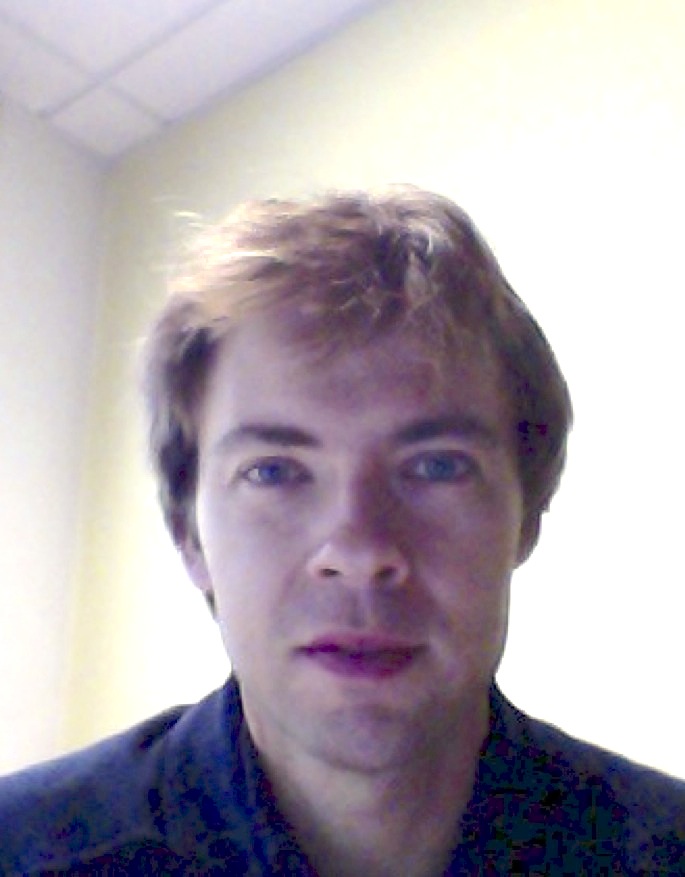}}]{Yezekael Hayel}
Yezekael Hayel received his M.Sc. in Computer Science and Applied
Mathematics from the University of Rennes 1, in 2003. He has a
Ph.D. in Computer Science from University of Rennes 1 and INRIA.
He is currently an assistant professor at the University of
Avignon, France. He works in the Networking group from 2006, this group is part of the computer science laboratory (LIA/CERI) of the University of Avignon. His research interests include game theoretical models applied to communication, social and transport networks. He is also interested in bio-inspired frameworks and pricing techniques for controlling and optimizing complex networks. He is involved in international projects and more information can be found at http://
http://lia.univ-avignon.fr/fileadmin/documents/Users/Intranet/chercheurs/hayel/acceuil.htm
\end{IEEEbiography}

\begin{IEEEbiography}[{\includegraphics[width=1in,height=1.25in,clip,keepaspectratio]{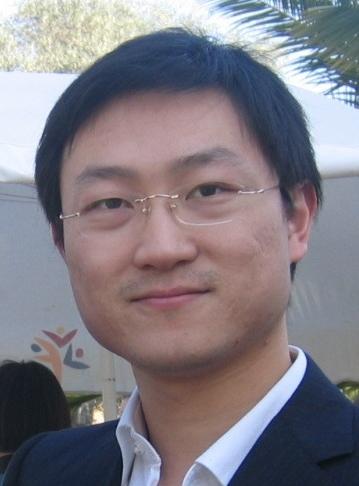}}]{Gaoning He}
Gaoning He entered the ƒcole d'IngŽnieurs TŽlŽcom ParisTech (France) in 2005 where he received his M.Sc and Ph.D. degrees respectively. He worked for Motorola Labs (Saclay, France) from 2006-2009 and Alcatel-Lucent Chair on Flexible Radio (Gif-sur-Yvette, France) until 2010. He then joined Alcatel-Lucent Bell-Labs in Shanghai until 2011. He is now working as a senior research engineer at Huawei Technologies Co. Ltd in Shanghai. His research interests include green wireless communications, information theory, applications of optimization theory and game theory.
\end{IEEEbiography}

\end{document}